\documentclass[rnote]{aa}

\usepackage{natbib,twoopt}
\usepackage[varg]{txfonts}
\usepackage[breaklinks=true]{hyperref} 
\usepackage{longtable}
\usepackage{graphicx}
\usepackage{txfonts}
\usepackage{multirow}
\usepackage{color}

\bibpunct{(}{)}{;}{a}{}{,}             
\makeatletter
  \newcommandtwoopt{\citeads}[3][][]{\href{http://adsabs.harvard.edu/abs/#3}%
    {\def\hyper@linkstart##1##2{}%
     \let\hyper@linkend\@empty\citealp[#1][#2]{#3}}}
  \newcommandtwoopt{\citepads}[3][][]{\href{http://adsabs.harvard.edu/abs/#3}%
    {\def\hyper@linkstart##1##2{}%
     \let\hyper@linkend\@empty\citep[#1][#2]{#3}}}
  \newcommandtwoopt{\citetads}[3][][]{\href{http://adsabs.harvard.edu/abs/#3}%
    {\def\hyper@linkstart##1##2{}%
     \let\hyper@linkend\@empty\citet[#1][#2]{#3}}}
  \newcommandtwoopt{\citeyearads}[3][][]%
    {\href{http://adsabs.harvard.edu/abs/#3}
    {\def\hyper@linkstart##1##2{}%
     \let\hyper@linkend\@empty\citeyear[#1][#2]{#3}}}
\makeatother

\def\Teff{\hbox{$\thinspace T_{\mathrm{eff}}$}}

\begin{document} 
\title{Masses and luminosities for 342 stars from the PennState-Toru\'n Centre for Astronomy Planet Search}

\author{M.~Adamczyk\and B.~Deka-Szymankiewicz\and A.~Niedzielski}

\institute{Toru\'n Centre for Astronomy, Nicolaus Copernicus University in Toru\'n, Grudziadzka 5, 87-100 Toru\'n, Poland, \email{Michalina.Gorecka@astri.umk.pl}\label{inst1}}

\date{Received date /
Accepted date }

\abstract {} {We present revised basic stellar astrophysical parameters:  masses, luminosities, ages and radii for 342 stars 
from PennState-Toru\'n Centre for Astronomy Planet Search. Atmospheric parameters for 327 stars are available 
from \citet{2012Zielinski}, for the remaining 15 objects we present also spectroscopic atmospheric parameters:  effective temperatures, 
surface gravities and iron abundances. } {Spectroscopic atmospheric parameters were obtained with a standard
 spectroscopic analysis procedure, using ARES \citep{2007Sousa} 
and MOOG \citep{1973Sneden} or TGVIT \citep{2005Takeda} codes. To refine stellar masses, ages and luminosities 
we applied a Bayesian method based on \citet{2005Jorgensen} formalism, modified by \citet{2006daSilva}.
}
{The revised stellar masses for 342 stars and their uncertainties are generally lower than those presented in \citet{2012Zielinski}.
Atmospheric parameters for 13 objects are determined here for the first time. }{}

\keywords{stars: low-mass -- stars: late-type -- stars: fundamental parameters -- stars: statistics}
\titlerunning{Bayesian masses for 342 PTPS stars.}

\maketitle

\section{Introduction}
Since the discovery of the first exoplanet through precise radial velocity measurements \citep{1995Mayor} 
the technique proved to be the most versatile and resulted in detection of over 430  extrasolar planetary 
systems around stars at various evolutionary stages known today. By nature, the planetary masses delivered 
by this technique are uncertain to the $\sin i$ factor, due to the unknown orbital plane inclination, and relative 
to the stellar mass. From the perspective of proper interpretation of the nature of the extrasolar planetary
 systems, precise determinations of the mass of their hosts are therefore invaluable.

One of the indirect approaches to obtain stellar mass is to compare available observational data like detailed 
spectroscopically determined atmospheric parameters, photometry or luminosities to theoretic stellar 
evolutionary models, for instance by isochrone fitting \citep{1982Flannery}. 

Given very approximate methodology applied in \citet{2012Zielinski} to derive stellar masses, we decided 
to revise previous results in a more sophisticated approach.
The purpose of this paper is to constrain better masses, ages, radii and, when no parallax is available also
luminosities, for red giants  presented in \citet{2012Zielinski} through Bayesian probability approach \citep{2006daSilva}
and to deliver for the first time atmospheric parameters, masses, ages and luminosities for the 16~stars omitted in that paper. 

The  stars considered here come from the Red Giant Clump subsample of the ongoing 
PennState-Toru\'n Centre for Astronomy Planet Search \citep[PTPS,][]{2007Niedzielski,2008Niedzielski}, 
a long-term project  focused on detection and characterization of planetary systems around stars at various evolutionary stages. 
Stellar atmospheric parameters together with uncertainties, required in further mass, age, luminosity determination, 
were taken from \citet{2012Zielinski} where effective temperatures, surface gravities, metallicities and microturbulence velocities were derived using TGVIT \citep{2005Takeda} from spectra obtained with the
Hobby-Eberly Telescope (HET, \citealt{1998Ramsey}) and its High Resolution Spectrograph (HRS,
\citealt{1998Tull}) operated in the queue scheduled mode \citep{2007Shetrone}. 

This paper is organized as follows: in Sect. \ref{sec2} we describe the sample and the method used in determination 
of stellar atmospheric parameters.  In Sect. \ref{sec3} mathematical formalism and example of constructed probability 
distribution function are provided. The results, a comparison with parameters obtained by \citet{2012Zielinski} 
and others is presented in Sect. \ref{sec4} while Sect. \ref{sec5} contains conclusions.

\section{The sample and atmospheric parameters}\label{sec2}

After we published the results of spectroscopic analysis for the first group of  stars observed in our project \citep{2012Zielinski}, 
eight of the objects were studied also by others. These studies revealed nearly perfect agreement 
in effective temperatures, $\log g$ and metallicities. Three objects (HD~102272, BD+20~2457 and BD+48~738) 
were studied by \citet{2013Mortier} who found our atmospheric parameters to agree within $1 \sigma$ with their, 
except [Fe/H] for BD+20 2457 that was found $\sim 3 \sigma$ lower. Another five objects 
(HD~17092, HD~240210, HD~240237, HD~96127 and HD~219415) were studied by \citet{2015Sousa}.
Again, very good agreement was found in all parameters. Given various input data and methods applied, 
we are confident that our atmospheric parameters are robust.

All available data on the program stars are summarized in \citet{2012Zielinski} where stellar masses were 
estimated by $\chi^2$ fitting  of stellar atmospheric parameters to \citet{2000Girardi} tracks for the nearest metallicity.

\subsection{Cross-correlation function analysis}

\begin{figure}
\begin{center}
  \includegraphics[scale=1.0]{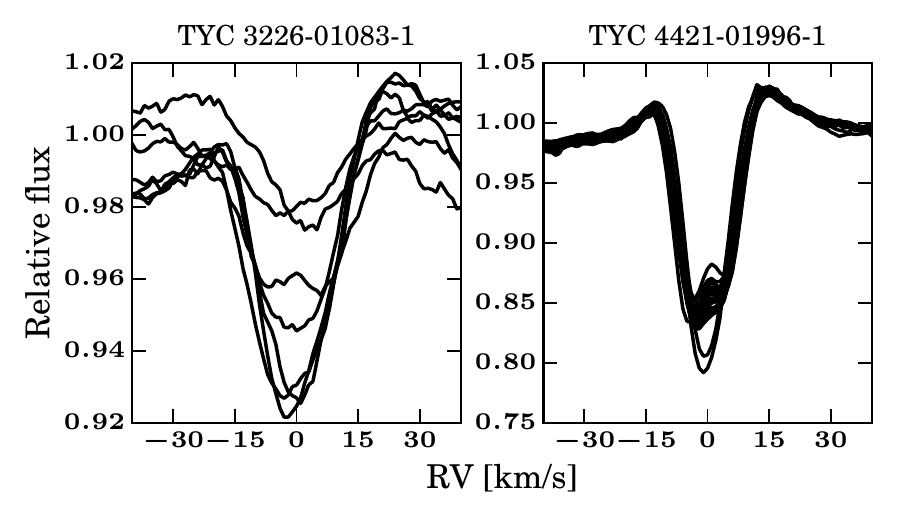}
  \caption{CCFs for TYC~3226-01083-1 and TYC~4421-01996-1. Left panel: weak and highly variable CCF,
               right panel: apparently unresolved spectroscopic binary.}
\label{fig:1}
\end{center}
\end{figure}

In search for the nature of the 16 stars for which spectroscopic analysis of \citet{2012Zielinski}
resulted in very uncertain data, we checked the complete sample discussed here for additional 
components and/or variable/peculiar cross-correlation functions \citep[CCF,][]{2012Nowak, 2013Nowak}. 
We found two objects which we consider multiple system due to variable CCF: TYC~0435-01209-1, TYC~4421-01996-1 (see Figure \ref{fig:1}).
We also found that TYC~3930-01790-1 and 3 out of 16 stars with incomplete data in \citet{2012Zielinski}:
TYC~0405-01700-1, TYC~3226-01083-1, TYC~3318-00020-1 are fast rotators with weak and variable CCF
(see Figure \ref{fig:1}), what is in good agreement with results of \citet{2014Adamow} for the last three objects. 
All other stars presented in \citet{2012Zielinski} show stable single CCF which makes them either single or SB1.

\subsection{Atmospheric parameters}

We adopted the atmospheric parameters  for 327 stars from paper by \citet{2012Zielinski}. 
Stars with variable or weak CCF identified in previous section were omitted in further analysis 
as their equivalent widths are very uncertain.

We also obtained atmospheric parameters for 13 stars (Table \ref{table:1}) from \citet{2012Zielinski},
for which these authors give only rough estimates of stellar parameters although our CCF analysis revealed a steady profile. 
We used ARES \citep{2007Sousa} 
to measure equivalent widths (EWs) for neutral (Fe~I) and ionized (Fe~II) iron absorption lines from the line list
of \citet{2005Takeda} as most suitable for our HET/HRS spectra \citep{2015Adamow}.
Next, we used the obtained EWs to determine atmospheric parameters with MOOG \citep{1973Sneden}.

Furthermore, with the procedure identical to that described in \citet{2012Zielinski} 
we obtained new atmospheric parameters for TYC~3011-00791-1{\footnote{The results for this star 
were already published in \citet{2015Niedzielski}} for which a better quality spectrum was available, 
and for TYC~4444-00200-1,  misidentified by \citet{2012Zielinski} (Table \ref{table:1a}).
 
The sample considered  here contains 342 relatively bright, field stars in total, mostly giants with 
 $\Teff$ between 4055~K and 6239~K (G8-K2 spectral type), $\log g$ between 1.39 and 4.78 and the [Fe/H]  between $-$1.0 to +0.45.

\begin{table}
\begin{center}
\caption{\label{table:1}Updated stellar atmospheric parameters for 13 objects from red clump PTPS sample stars.}
\begin{tabular}{crcc}
\hline
Name TYC & [Fe/H] & $\Teff$ [K]$^{*}$&  $\log g^{\dagger}$\\
\hline
0405-00684-1& $-$0.11$\pm$ 0.14& 4750& 1.5 \\
1062-00017-1& $-$0.07$\pm$ 0.13& 5000& 2.0 \\ 
1496-01016-1& $-$0.81$\pm$ 0.10& 4750& 3.0 \\ 
1496-00374-1&  0.00$\pm$ 0.23& 4500& 2.5 \\ 
2818-00602-1& $-$0.45$\pm$ 0.17& 4750& 1.5 \\ 
2818-00990-1& $-$0.07$\pm$ 0.12& 5250& 1.5 \\   
3020-01288-1& $-$0.17$\pm$ 0.14& 4500& 2.0 \\ 
3105-01103-1& $-$0.03$\pm$ 0.16& 4500& 1.5 \\ 
3226-00868-1& $-$0.51$\pm$ 0.73& 4250& 3.0 \\
3304-00479-1& $-$0.18$\pm$ 0.12& 4750& 1.5 \\
3431-00086-1& $-$0.27$\pm$ 0.26& 4750& 1.5 \\ 
3663-00838-1& $-$0.06$\pm$ 0.30& 5000& 2.0 \\ 
4006-00890-1&  0.21$\pm$ 0.21& 4000& 2.0 \\
\hline
\multicolumn{4}{l}{$^{*}$\footnotesize{All values with uncertainty 250~K}}\\ 
\multicolumn{4}{l}{$^{\dagger}$\footnotesize{All values with uncertainty 0.5~dex}}\\ 
\end{tabular}
\end{center}
\end{table}

\begin{table}
\begin{center}
\caption{\label{table:1a}New stellar atmospheric parameters for 2 misidentified objects from red clump PTPS sample stars.}
\begin{tabular}{crcc}
\hline
Name TYC & [Fe/H] & $\Teff$ [K]&  $\log g$\\
\hline
3011-00791-1& $-$0.18$\pm$0.05& 4218$\pm$69 & 1.78$\pm$0.30 \\
4444-00200-1&  0.17$\pm$0.03& 4792$\pm$60 & 3.27$\pm$0.18 \\
\hline
\end{tabular}
\end{center}
\end{table}

\section{Construction of the probability distribution function}\label{sec3}

For the Bayesian analysis we adopted theoretical stellar models from \citet{2012Bressan} gathered from
the interactive interface on Osservatorio Astronomico di Padova website (http:\/\/stev.oapd.inaf.it\/cgi-bin\/cmd).
We used isochrones with metallicity $Z=0.0001$, $0.0004$, $0.0008$, $0.001$, $0.002$, $0.004$, $0.006$, $0.008$, $0.01$, $0.0152$,
$0.02$, $0.025$, $0.03$, $0.04$, $0.05$, $0.06$ and $0.008$ interval in $\log \mathrm{(age/yr)}$. The adopted solar distribution of heavy elements 
corresponds to Sun's metallicity: $Z \simeq 0.0152$ \citep{2011Caffau}. 
The helium abundance for a given metallicity was obtained from relation $Y=0.2485+1.78~Z$. 

\subsection{Mathematical formalism}
We implemented the Bayesian method based on \citet{2005Jorgensen} formalism and modified by \citet{2006daSilva} 
to avoid statistical biases and to take uncertainty estimates of  observed quantities into consideration. 
For a given star, represented by full set of available atmospheric parameters
(and luminosity if parallax was available):  ($\mathrm{[Fe/H]} \pm \sigma_{\mathrm{[Fe/H]}}$, 
$\log \Teff \pm \sigma_{\log \Teff}$, 
$\log g \pm \sigma_{\log g}$,
$\log L \pm \sigma_{\log L}$), 
isochrone of [Fe/H] and age $t$ we calculated 
the probability of belonging to a given mass range.

The Initial Mass function for single star was taken from \citet{1955Salpeter}. Instead of absolute magnitude
like \citet{2006daSilva} we used luminosity (if parallax was available) and logarithm of surface gravity. 
We further followed the procedure detailed in \citet{2006daSilva} and calculated searched quantities 
({\it e.g.} mass, luminosity and age) and their uncertainties  from basic parameters (mean, variance) 
of the normalized probability distribution functions (PDFs). 
 
For stars with Hipparcos \citep{2007vanLeeuwen} parallaxes, for which  $\pi > \sigma_{\pi}$,
stellar luminosities were calculated directly and only stellar mass and age were obtained from PDFs.

\subsection{Stellar radii} 
With either stellar mass or luminosity derived from the Bayesian analysis and available 
atmospheric parameters one can calculate stellar radii as either:
\begin{equation}
\label{eq:4}
R/R_{\odot}(\Teff,L) = \left(\frac{L}{L_{\odot}}\right)^{1/2}\left(\frac{\Teff_{\odot}}{\Teff}\right)^2,
\end{equation}
or
\begin{equation}
\label{eq:5}
R/R_{\odot}(g,M) = \left(\frac{M}{M_{\odot}}\frac{g_{\odot}}{g}\right)^{1/2},
\end{equation}
where $R_{\odot}, M_{\odot}, L_{\odot}, \Teff_{\odot}$ are solar values of radius, mass, luminosity
and effective temperature. We adopted the mean value of those two determinations 
and the total derivative as the  radius uncertainty estimate.

\section{Results}\label{sec4}
A typical examples of PDFs for luminosity, age and mass with apparently unique set of solutions 
are presented in Figure \ref{fig:2}. In comparison with other parameters, the widest distribution and 
the lowest probability peak is present in the case of $\log \mathrm{(age/yr)}$. As a consequence stellar ages obtained here are most uncertain.
For 11 objects the PDF shows double peaks (see Figure \ref{fig:3}). For such stars we estimated 
two separate sets of solutions, that with higher value of PDF for stellar mass is listed first. The reason 
for this ambiguity is higher density (degeneracy) of stellar evolutionary tracks in the part 
of Hertzsprung-Russell diagram occupied by many of our stars. The probability of obtaining
non-unique and well separated solutions is higher for Red Giant Clump stars, common in our sample.

The resulting stellar parameters with their uncertainties are presented in Table \ref{table:2}, 
where we also present average values of radii calculated from Eq. \ref{eq:4} and \ref{eq:5}.

\begin{figure}
\begin{center}
\includegraphics[scale=0.9]{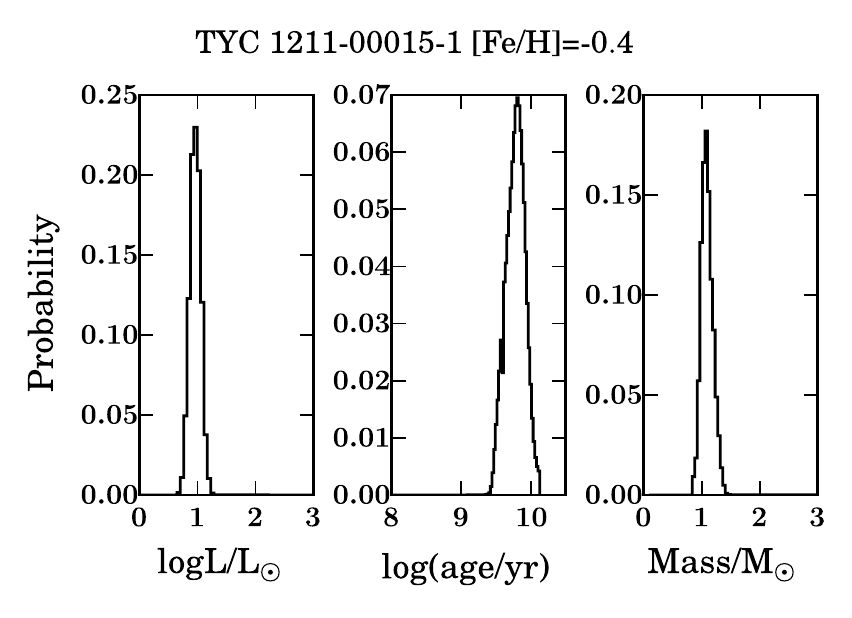}
\caption{Probability distribution functions for luminosity, mass and age of TYC~1211-00015-1.}
\label{fig:2}
\end{center}
\end{figure}

\begin{figure}
\begin{center}
\includegraphics[scale=.9]{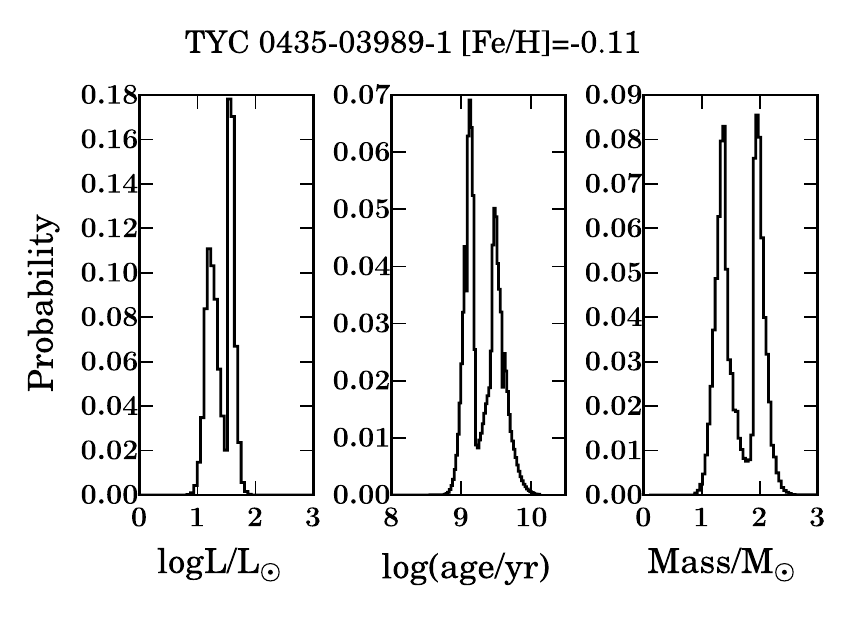}
\caption{Double peak probability distribution functions for luminosity, mass and age of TYC~0435-03989-1.}
\label{fig:3}
\end{center}
\end{figure}

The obtained stellar masses range between 0.88 and 3.75 solar masses with vast majority of stars (240 stars)
between 1.0 and $1.5 M_{\odot}$. The mean uncertainty in stellar mass is $0.19 M_{\odot}$.   
For  $\log L/L_{\odot}$ we obtained values  between 0.04 and 3.1 with 164 objects with luminosities
between $\log L/L_{\odot}$ 1.5 and 2.0. For most of stars the uncertainty in $\log L/L_{\odot}$ is lower
than 0.3, for 288 stars with no trigonometric parallaxes the mean uncertainty in $\log L/L_{\odot}$ is 0.13. 

Stellar ages, $\log \mathrm{(age/yr)}$, range from 8.36 to 10.09 with mean uncertainty of $\sim 0.18$~dex.
 
The radii determined from Eq. \ref{eq:4} differ in most of cases from those derived from Eq. \ref{eq:5} 
by more than the estimated uncertainties. In most cases (54\%) the difference is lower than 
10\% of the mean value, for 80\% of stars it stays within 30\% but uncertainties as high as 100\% may happen.
In Table \ref{table:2} we present mean values of radii derived from both equations.
The resulting radii range from 1.32 to 50.24 solar radius with mean uncertainty of 3.16 solar radius.

\subsection{Comparison with \citet{2012Zielinski}}
For 288 objects without trigonometric parallax, luminosities available derived by \citet{2012Zielinski} agree 
with ours very well with Pearsons correlation coefficient of $r=0.80$. Our determinations are, however, 
more precise. Estimated uncertainties are nearly two times lower than those of \citet{2012Zielinski}.
As a consequence, for these stars the radii calculated from Eq. \ref{eq:4} are also in 
good agreement. 
Our $\log \mathrm{(age/yr)}$ determinations are in general larger by 0.15 dex, with $r=0.73$\footnote{To compare 
with \citet{2012Zielinski}, when given range in $\log \mathrm{(age/yr)}$, the maximum value was taken.}.

Stellar masses obtained here, with average value of $1.35 M_{\odot}$, are generally lower 
by over $0.1 M_{\odot}$ than those of \citet{2012Zielinski} - $1.5 M_{\odot}$, but at the same time 
the abundant population of stars with masses below solar is now absent. The resulting stellar masses 
are also more precise as the average uncertainty in \citet{2012Zielinski} is $0.28 M_{\odot}$. 
The comparison of stellar masses as well as other parameters derived here is compared 
with a that of \citet{2012Zielinski} in Figure \ref{fig:4}. 

\begin{figure}
\begin{center}
\includegraphics[scale=1.0]{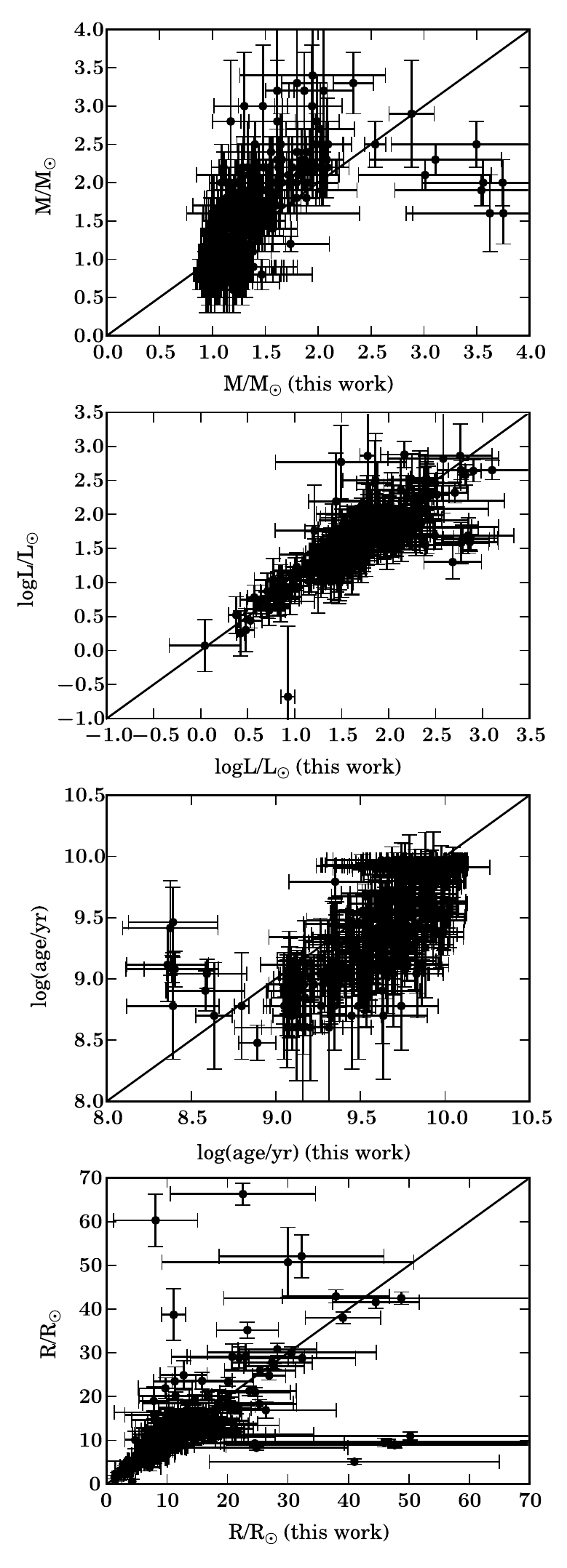}
\caption{Comparison of masses, luminosities, ages and radii with their uncertainties from this work and from \citet{2012Zielinski}.}
\label{fig:4}
\end{center}
\end{figure}

\subsection{Comparison with other determinations}

For eight stars our stellar masses can be compared with determinations by other authors.
For HD~102272, BD+20~2457 and BD+48~740 we note good agreement with \citet{2013Mortier}, 
well within $1 \sigma$ uncertainty. Out of five stars for which stellar masses were obtained by \citet{2015Sousa} 
for all but one we have good agreement within $1 \sigma$ as well. The only exception is HD~240237 
for which our new estimated stellar mass, $1.46 \pm 0.32 M_{\odot}$  agrees within estimated uncertainties 
with those presented in \citet{2012Gettel} and \citet{2012Zielinski} while \citet{2015Sousa} found for this
star a mass of $0.614 \pm 0.076 M_{\odot}$. We find their result rather uncertain as it is hard to believe
that a star of that mass is already a giant with $\log g=1.66\pm0.15$. Indeed,  for the stellar mass of \citet{2015Sousa} the web interface
for the Bayesian estimation of stellar parameters{\footnote{http://stev.oapd.inaf.it/cgi-bin/param}}
returns $\log g=4.662\pm 0.358$ for this object, contrary to what these authors present. 

\subsection{Comparison with luminosities from trigonometric parallaxes}
For 54 objects with trigonometric parallaxes we compared our Bayesian luminosities estimates 
(using only $\log \Teff$, $\log g$ and [Fe/H] with their uncertainties as a stellar parameter input)
with those calculated directly from trigonometric parallaxes (Figure \ref{fig:5}) and 
we found a relation $\log L/L_{\odot,\pi} = (0.68\pm0.11)\log L/L_{\odot} + (0.72\pm0.18)$ and the Pearson 
correlation coefficient of $r=0.66$. As  expected the highest scatter is present for stars with lowest parallaxes $\pi < 5$~mas.
It is obvious that the GAIA \citep{2001Perryman} will provide parallaxes with high accuracy 
which will allow to determine better  luminosities and in a consequence stellar masses for the stars presented here.

\begin{figure}
\begin{center}
\includegraphics[scale=1.0]{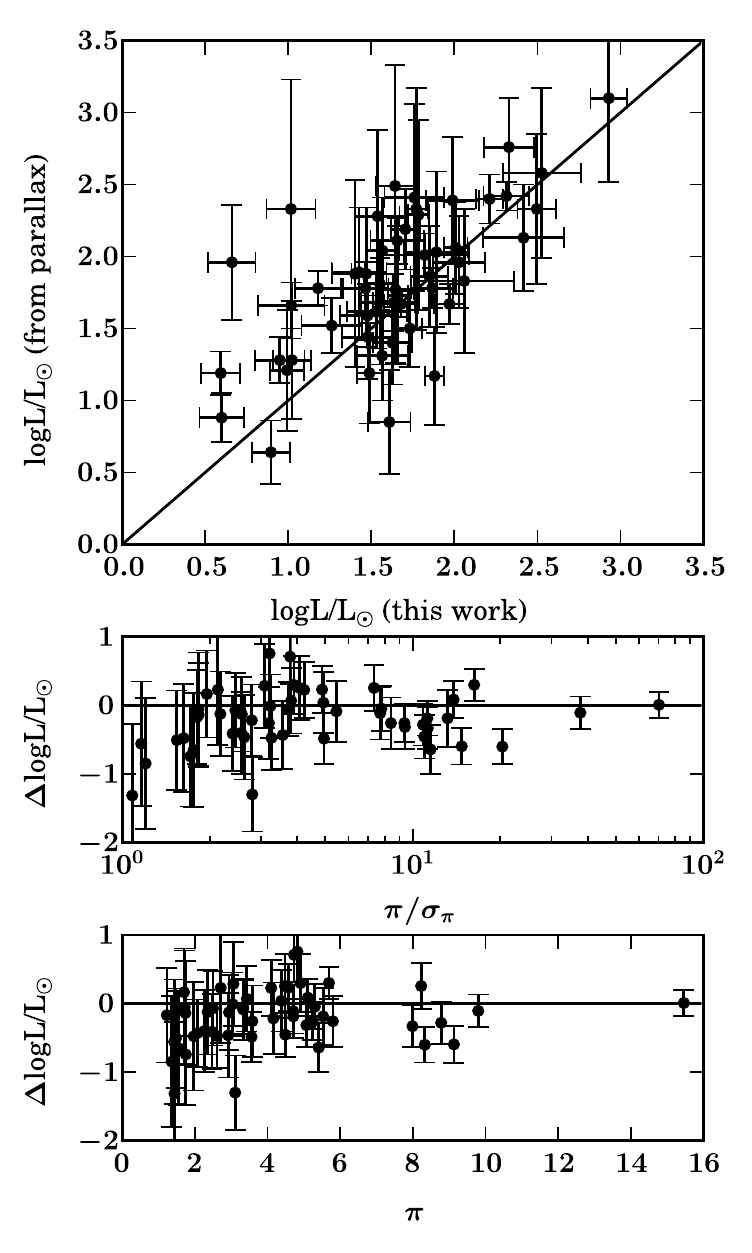}
\caption{Comparison of luminosities calculated from trigonometric parallaxes with our Bayesian estimates.}
\label{fig:5}
\end{center}
\end{figure}

\section{Conclusions}\label{sec5}
We presented stellar masses, luminosities and $\log$(age/yr), obtained through a  Bayesian analysis 
of atmospheric parameters available from \citet{2012Zielinski} as well as 
new estimates of radii for 342 stars, targets of the PTPS.  For 13 stars we present atmospheric parameters for the first time, for another two we updated ones.
The results, as based on specific set of stellar evolutionary models are obviously model - dependent. 
  Based on results of \citet{2014Thompson} we expect, however,  that in the mass range occupied by our targets this introduces inaccuracy  below estimated uncertainties.

As a consequence of the adopted stellar models our stellar masses take into consideration the mass-loss
at the RGB but the resulting masses represent ZAMS masses for all stars. This obvious simplification 
is well justified by the stellar mass range we are dealing with in our project. It is not expected that 
a solar-mass star loses more than $0.09\pm 0.03 M_{\odot}$ during its evolution up to 
the Horizontal Branch \citep{2012Miglio} which effect contributes only to  a minor fraction of estimated uncertainties.

Presented here stellar masses represent an important improvement over previous determination 
due to application of much denser set of stellar models, and more detailed treatment of metallicities. 
As a result,  significant uncertainty decrease in stellar masses was achieved. An important result
is also throughout test of an improved tool to derive stellar masses from available stellar atmospheric 
parameters in the PTPS sample.

\begin{acknowledgements}
We acknowledge the financial support from the Polish National Science Centre
through a grant 2012/07/B/ST9/04415.  

The HET is a joint project of the University of Texas at Austin, the Pennsylvania State
University, Stanford University, Ludwig- Maximilians-Universit\"at M\"unchen,
and Georg-August-Universit\"at G\"ottingen. The HET is named in honor of its
principal benefactors, William P. Hobby and Robert E. Eberly. 

\end{acknowledgements}

\bibliographystyle{aa} 
\bibliography{aa} 

\onecolumn
\begin{longtab}
\begin{longtable}{lllccccc}
\caption{\label{table:2}Recalculated astrophysics stellar parameters for the red clump PTPS sample stars.}\\
\hline\hline
\multicolumn{3}{c}{Names} & \multirow{2}{*}{$M/M_{\odot}$} & \multirow{2}{*}{$\log\mathrm{(age/yr)}$} & \multirow{2}{*}{$\log L/L_{\odot}$} & \multirow{2}{*}{$\log L/L_{\odot,\pi}$}  &  \multirow{2}{*}{$R/R_{\odot}$}\\
TYC & HD & BD & & & & & \\
\hline
\endfirsthead
\caption{continued.}\\
\hline\hline
\multicolumn{3}{c}{Names} & \multirow{2}{*}{$M/M_{\odot}$} & \multirow{2}{*}{$\log\mathrm{(age/yr)}$} & \multirow{2}{*}{$\log L/L_{\odot}$} & \multirow{2}{*}{$\log L/L_{\odot,\pi}$}  &  \multirow{2}{*}{$R/R_{\odot}$}\\
TYC & HD & BD & & & & & \\
\hline
\endhead
\hline
\endfoot
0014-00693-1 & - & +04 126 & 1.42 $\pm$ 0.23 & 9.44 $\pm$ 0.21 & 1.78 $\pm$ 0.13 & - & 11.32 $\pm$ 2.14 \\
0014-00731-1 & 4446 & +04 118 & 1.32 $\pm$ 0.18 & 9.55 $\pm$ 0.19 & 1.65 $\pm$ 0.11 & 1.77 $\pm$ 0.51 & 10.45 $\pm$ 4.18 \\
0014-00752-1 & - & +04 119 & 1.23 $\pm$ 0.22 & 9.66 $\pm$ 0.24 & 1.74 $\pm$ 0.12 & - & 11.16 $\pm$ 2.10 \\
0014-00769-1 & - & +04 107 & 0.93 $\pm$ 0.05 & 10.04 $\pm$ 0.09 & 1.84 $\pm$ 0.05 & - & 12.01 $\pm$ 1.10 \\
0014-00882-1 & - & +04 112 & 1.21 $\pm$ 0.20 & 9.79 $\pm$ 0.23 & 1.90 $\pm$ 0.12 & - & 15.05 $\pm$ 2.77 \\
0017-00572-1 & - & +04 122 & 0.95 $\pm$ 0.06 & 10.04 $\pm$ 0.09 & 1.66 $\pm$ 0.06 & 1.73 $\pm$ 0.48 & 10.39 $\pm$ 3.92 \\
0017-00668-1 & - & - & 1.11 $\pm$ 0.18 & 9.80 $\pm$ 0.23 & 1.54 $\pm$ 0.14 & - & 8.27 $\pm$ 1.78 \\
0017-00900-1 & - & +05 107 & 0.96 $\pm$ 0.07 & 10.02 $\pm$ 0.11 & 1.74 $\pm$ 0.06 & - & 11.07 $\pm$ 1.16 \\
0017-01084-1 & 4760 & +05 109 & 1.06 $\pm$ 0.19 & 9.76 $\pm$ 0.24 & 2.93 $\pm$ 0.11 & 3.10 $\pm$ 0.58 & 48.75 $\pm$ 29.36 \\
0017-01136-1 & - & +04 111 & 1.28 $\pm$ 0.20 & 9.57 $\pm$ 0.21 & 1.77 $\pm$ 0.13 & - & 11.37 $\pm$ 2.13 \\
0017-01292-1 & - & +04 114 & 1.11 $\pm$ 0.12 & 9.89 $\pm$ 0.16 & 0.53 $\pm$ 0.13 & - & 2.37 $\pm$ 0.50 \\
0017-01299-1 & 4366 & +04 113 & 1.07 $\pm$ 0.13 & 9.95 $\pm$ 0.17 & 1.30 $\pm$ 0.12 & - & 6.19 $\pm$ 1.28 \\
0018-00446-1 & - & +04 132 & 1.24 $\pm$ 0.15 & 9.66 $\pm$ 0.18 & 1.52 $\pm$ 0.11 & - & 8.45 $\pm$ 1.24 \\
0096-00005-1 & - & +06 758 & 1.29 $\pm$ 0.30 & 9.75 $\pm$ 0.28 & 1.95 $\pm$ 0.19 & - & 13.78 $\pm$ 4.72 \\
0096-00109-1 & - & +06 750 & 2.09 $\pm$ 0.14 & 9.05 $\pm$ 0.08 & 1.64 $\pm$ 0.06 & - & 8.59 $\pm$ 0.94 \\
0096-00163-1 & - & - & 1.21 $\pm$ 0.20 & 9.75 $\pm$ 0.22 & 1.41 $\pm$ 0.15 & - & 7.47 $\pm$ 1.62 \\
0096-00301-1 & - & - & 1.35 $\pm$ 0.17 & 9.60 $\pm$ 0.18 & 1.48 $\pm$ 0.13 & - & 8.27 $\pm$ 1.50 \\
0096-00371-1 & - & +06 755 & 1.17 $\pm$ 0.16 & 9.84 $\pm$ 0.19 & 1.49 $\pm$ 0.11 & - & 8.42 $\pm$ 1.58 \\
0096-00378-1 & - & - & 1.02 $\pm$ 0.10 & 9.90 $\pm$ 0.15 & 1.57 $\pm$ 0.08 & - & 8.91 $\pm$ 0.97 \\
0096-00417-1 & - & - & 1.57 $\pm$ 0.23 & 9.35 $\pm$ 0.18 & 1.83 $\pm$ 0.12 & - & 12.55 $\pm$ 2.03 \\
0096-00418-1 & - & - & 1.01 $\pm$ 0.10 & 9.97 $\pm$ 0.14 & 1.52 $\pm$ 0.07 & - & 8.58 $\pm$ 1.05 \\
0096-00659-1 & 30897 & +05 751 & 1.41 $\pm$ 0.23 & 9.62 $\pm$ 0.21 & 0.66 $\pm$ 0.14 & 1.96 $\pm$ 0.40 & 8.81 $\pm$ 4.00 \\
0096-00708-1 & - & - & 2.09 $\pm$ 0.15 & 9.07 $\pm$ 0.09 & 1.65 $\pm$ 0.08 & - & 8.89 $\pm$ 1.23 \\
0096-00732-1 & - & - & 1.15 $\pm$ 0.09 & 9.74 $\pm$ 0.11 & 0.83 $\pm$ 0.07 & - & 3.38 $\pm$ 0.32 \\
0096-00778-1 & - & +06 754 & 1.02 $\pm$ 0.11 & 9.96 $\pm$ 0.15 & 1.33 $\pm$ 0.10 & - & 6.57 $\pm$ 1.03 \\
0096-00887-1 & - & - & 1.17 $\pm$ 0.17 & 9.81 $\pm$ 0.20 & 1.44 $\pm$ 0.12 & - & 7.89 $\pm$ 1.41 \\
0272-00909-1 & - & - & 1.04 $\pm$ 0.07 & 10.03 $\pm$ 0.09 & 1.42 $\pm$ 0.06 & - & 7.55 $\pm$ 0.89 \\
0273-00125-1 & - & +01 2626 & 1.10 $\pm$ 0.21 & 9.81 $\pm$ 0.25 & 2.29 $\pm$ 0.13 & - & 20.05 $\pm$ 4.97 \\
0273-00150-1 & - & - & 1.03 $\pm$ 0.07 & 10.00 $\pm$ 0.10 & 0.64 $\pm$ 0.08 & - & 2.90 $\pm$ 0.35 \\
0273-00224-1 & 103485 & +02 2493 & 1.11 $\pm$ 0.21 & 9.79 $\pm$ 0.25 & 2.51 $\pm$ 0.13 & - & 27.37 $\pm$ 6.69 \\
0273-00279-1 & - & +02 2492 & 1.00 $\pm$ 0.07 & 10.03 $\pm$ 0.10 & 0.48 $\pm$ 0.09 & - & 2.14 $\pm$ 0.38 \\
0273-00451-1 & - & - & 1.46 $\pm$ 0.23 & 9.36 $\pm$ 0.20 & 1.83 $\pm$ 0.14 & - & 11.56 $\pm$ 2.30 \\
0273-00669-1 & 102842 & +01 2619 & 1.40 $\pm$ 0.27 & 9.43 $\pm$ 0.25 & 1.87 $\pm$ 0.16 & - & 12.35 $\pm$ 2.93 \\
0276-00507-1 & - & +03 2562 & 1.14 $\pm$ 0.25 & 9.72 $\pm$ 0.28 & 2.70 $\pm$ 0.14 & - & 32.35 $\pm$ 8.82 \\
0400-00329-1 & - & - & 1.40 $\pm$ 0.17 & 9.49 $\pm$ 0.17 & 1.64 $\pm$ 0.10 & - & 9.75 $\pm$ 1.42 \\
0401-01176-1 & - & - & 1.00 $\pm$ 0.06 & 9.98 $\pm$ 0.09 & 0.87 $\pm$ 0.08 & - & 3.72 $\pm$ 0.44 \\
0401-01874-1 & - & +01 3439 & 1.42 $\pm$ 0.24 & 9.46 $\pm$ 0.21 & 1.68 $\pm$ 0.15 & - & 9.58 $\pm$ 1.94 \\
0401-02049-1 & 157855 & +01 3432 & 1.04 $\pm$ 0.08 & 10.00 $\pm$ 0.11 & 1.49 $\pm$ 0.07 & 1.19 $\pm$ 0.35 & 7.06 $\pm$ 1.99 \\
0401-02075-1 & - & - & 1.15 $\pm$ 0.15 & 9.78 $\pm$ 0.18 & 1.62 $\pm$ 0.09 & - & 9.86 $\pm$ 1.40 \\
0405-00236-1 & - & +02 3328 & 1.14 $\pm$ 0.18 & 9.74 $\pm$ 0.21 & 1.73 $\pm$ 0.11 & - & 10.86 $\pm$ 1.94 \\
0405-00405-1 & - & +02 3317 & 1.89 $\pm$ 0.14 & 9.05 $\pm$ 0.09 & 1.96 $\pm$ 0.06 & - & 13.32 $\pm$ 1.22 \\
0405-00414-1 & - & - & 1.03 $\pm$ 0.13 & 9.93 $\pm$ 0.18 & 1.45 $\pm$ 0.12 & - & 7.31 $\pm$ 1.37 \\
0405-00510-1 & - & +03 3401 & 1.17 $\pm$ 0.16 & 9.76 $\pm$ 0.20 & 1.60 $\pm$ 0.11 & - & 9.49 $\pm$ 1.65 \\
0405-00581-1 & - & +02 3322 & 1.13 $\pm$ 0.12 & 9.90 $\pm$ 0.15 & 1.49 $\pm$ 0.08 & - & 8.93 $\pm$ 1.13 \\
0405-00684-1 & - & +02 3308 & 3.74 $\pm$ 0.76 & 8.36 $\pm$ 0.24 & 2.78 $\pm$ 0.29 & - & 46.65 $\pm$ 27.24 \\
0405-01114-1 & - & +02 3313 & 1.11 $\pm$ 0.13 & 9.95 $\pm$ 0.17 & 1.54 $\pm$ 0.11 & - & 9.17 $\pm$ 1.69 \\
0405-01633-1 & 157936A & +02 3314A & 1.02 $\pm$ 0.11 & 9.96 $\pm$ 0.15 & 1.71 $\pm$ 0.08 & - & 10.69 $\pm$ 1.58 \\
0405-01634-1 & - & - & 1.39 $\pm$ 0.16 & 9.50 $\pm$ 0.15 & 1.59 $\pm$ 0.10 & - & 9.16 $\pm$ 1.32 \\
0405-01855-1 & 158253 & +02 3321 & 1.20 $\pm$ 0.24 & 9.74 $\pm$ 0.26 & 2.34 $\pm$ 0.13 & - & 25.40 $\pm$ 5.88 \\
0418-00640-1 & - & - & 0.88 $\pm$ 0.03 & 10.08 $\pm$ 0.05 & 2.30 $\pm$ 0.05 & - & 20.10 $\pm$ 1.71 \\
0418-00912-1 & - & +02 3332 & 1.21 $\pm$ 0.17 & 9.68 $\pm$ 0.20 & 1.57 $\pm$ 0.12 & - & 8.87 $\pm$ 1.40 \\
0430-00356-1 & - & +01 3567 & 1.38 $\pm$ 0.19 & 9.57 $\pm$ 0.19 & 1.60 $\pm$ 0.11 & - & 9.70 $\pm$ 1.70 \\
0430-02902-1 & 165592 & +01 3593 & 1.52 $\pm$ 0.25 & 9.37 $\pm$ 0.20 & 1.70 $\pm$ 0.16 & - & 9.52 $\pm$ 1.95 \\
0434-00031-1 & - & +02 3465 & 1.08 $\pm$ 0.13 & 9.93 $\pm$ 0.17 & 1.59 $\pm$ 0.11 & - & 9.87 $\pm$ 1.64 \\
0434-00551-1 & - & - & 1.39 $\pm$ 0.26 & 9.53 $\pm$ 0.24 & 1.77 $\pm$ 0.08 & - & 10.97 $\pm$ 1.65 \\
0434-00632-1 & 164734 & +03 3572 & 1.99 $\pm$ 0.21 & 9.10 $\pm$ 0.11 & 1.73 $\pm$ 0.07 & 1.50 $\pm$ 0.27 & 8.40 $\pm$ 1.92 \\
0434-00756-1 & - & +03 3584 & 1.84 $\pm$ 0.18 & 9.21 $\pm$ 0.11 & 1.61 $\pm$ 0.12 & - & 8.54 $\pm$ 1.43 \\
0434-01143-1 & - & - & 1.61 $\pm$ 0.19 & 9.36 $\pm$ 0.14 & 1.72 $\pm$ 0.09 & - & 10.31 $\pm$ 1.44 \\
0434-01505-1 & - & +03 3590 & 1.22 $\pm$ 0.13 & 9.67 $\pm$ 0.16 & 1.63 $\pm$ 0.08 & - & 9.73 $\pm$ 1.09 \\
0434-03595-1 & 165574 & +02 3489 & 1.27 $\pm$ 0.29 & 9.69 $\pm$ 0.29 & 1.76 $\pm$ 0.18 & 2.41 $\pm$ 0.65 & 18.43 $\pm$ 12.43 \\
0434-03897-1 & - & - & 1.22 $\pm$ 0.18 & 9.69 $\pm$ 0.22 & 1.54 $\pm$ 0.13 & - & 8.67 $\pm$ 1.62 \\
0434-04234-1 & 165742 & +02 3493 & 1.12 $\pm$ 0.15 & 9.83 $\pm$ 0.19 & 2.21 $\pm$ 0.08 & 2.40 $\pm$ 0.17 & 23.29 $\pm$ 5.12 \\
0434-04538-1 & - & - & 1.04 $\pm$ 0.14 & 9.88 $\pm$ 0.19 & 1.67 $\pm$ 0.09 & - & 9.99 $\pm$ 1.60 \\
0434-04779-1 & 165419 & +01 3585 & 1.12 $\pm$ 0.16 & 9.89 $\pm$ 0.20 & 1.78 $\pm$ 0.10 & - & 12.27 $\pm$ 2.25 \\
0435-03332-1 & - & +02 3497 & 1.81 $\pm$ 0.24 & 9.22 $\pm$ 0.17 & 1.47 $\pm$ 0.15 & - & 7.08 $\pm$ 1.23 \\
0683-00667-1 & - & +07 735 & 1.61 $\pm$ 0.11 & 9.31 $\pm$ 0.08 & 1.80 $\pm$ 0.03 & - & 11.16 $\pm$ 1.04 \\
0683-00789-1 & - & +07 721 & 1.36 $\pm$ 0.20 & 9.53 $\pm$ 0.20 & 1.67 $\pm$ 0.12 & - & 10.10 $\pm$ 1.83 \\
0683-01063-1 & - & - & 1.25 $\pm$ 0.05 & 9.31 $\pm$ 0.13 & 0.38 $\pm$ 0.08 & - & 1.32 $\pm$ 0.21 \\
0683-01190-1 & - & +07 736 & 1.09 $\pm$ 0.11 & 9.86 $\pm$ 0.16 & 1.65 $\pm$ 0.10 & - & 10.64 $\pm$ 2.32 \\
0684-00553-1 & - & - & 1.61 $\pm$ 0.34 & 9.35 $\pm$ 0.29 & 1.90 $\pm$ 0.14 & - & 13.42 $\pm$ 2.57 \\
0684-00744-1 & - & +07 742 & 1.24 $\pm$ 0.26 & 9.75 $\pm$ 0.26 & 2.10 $\pm$ 0.15 & - & 18.56 $\pm$ 4.85 \\
0684-01276-1 & - & - & 1.30 $\pm$ 0.15 & 9.66 $\pm$ 0.17 & 1.50 $\pm$ 0.10 & - & 8.58 $\pm$ 1.24 \\
0863-00082-1 & - & +15 2372 & 1.26 $\pm$ 0.19 & 9.63 $\pm$ 0.21 & 1.30 $\pm$ 0.15 & - & 6.29 $\pm$ 1.39 \\
0863-00230-1 & - & +15 2371 & 1.35 $\pm$ 0.25 & 9.63 $\pm$ 0.24 & 1.37 $\pm$ 0.21 & - & 7.13 $\pm$ 1.99 \\
0870-00084-1 & - & - & 1.06 $\pm$ 0.14 & 9.88 $\pm$ 0.19 & 0.82 $\pm$ 0.18 & - & 3.26 $\pm$ 0.99 \\
0870-00114-1 & 102272 & +14 2434 & 1.01 $\pm$ 0.12 & 9.91 $\pm$ 0.17 & 1.63 $\pm$ 0.12 & 1.40 $\pm$ 0.29 & 8.02 $\pm$ 2.14 \\
0870-00130-1 & - & +15 2387 & 0.99 $\pm$ 0.09 & 9.94 $\pm$ 0.14 & 0.99 $\pm$ 0.10 & 1.21 $\pm$ 0.42 & 4.73 $\pm$ 1.71 \\
0870-00204-1 & - & +15 2375 & 1.08 $\pm$ 0.14 & 9.87 $\pm$ 0.18 & 1.57 $\pm$ 0.10 & - & 8.95 $\pm$ 1.47 \\
0870-00207-1 & - & +15 2386 & 1.01 $\pm$ 0.16 & 9.88 $\pm$ 0.21 & 2.22 $\pm$ 0.11 & - & 17.22 $\pm$ 3.59 \\
0870-00241-1 & 102103 & +15 2374 & 1.56 $\pm$ 0.28 & 9.42 $\pm$ 0.22 & 1.82 $\pm$ 0.26 & 2.01 $\pm$ 0.15 & 14.52 $\pm$ 4.61 \\
0870-00255-1 & - & +15 2390 & 1.13 $\pm$ 0.16 & 9.85 $\pm$ 0.19 & 0.99 $\pm$ 0.16 & - & 4.36 $\pm$ 1.15 \\
0870-00314-1 & - & - & 1.13 $\pm$ 0.15 & 9.88 $\pm$ 0.19 & 1.01 $\pm$ 0.17 & - & 4.47 $\pm$ 1.25 \\
0870-00937-1 & - & +15 2385 & 1.26 $\pm$ 0.22 & 9.71 $\pm$ 0.23 & 1.27 $\pm$ 0.18 & - & 6.31 $\pm$ 1.63 \\
0955-00126-1 & - & - & 0.98 $\pm$ 0.06 & 10.00 $\pm$ 0.09 & 0.70 $\pm$ 0.10 & - & 2.99 $\pm$ 0.44 \\
0956-00028-1 & - & +14 2970 & 0.97 $\pm$ 0.15 & 9.88 $\pm$ 0.20 & 2.90 $\pm$ 0.09 & - & 44.52 $\pm$ 7.16 \\
1058-00120-1 & 188105 & +07 4275 & 1.17 $\pm$ 0.19 & 9.78 $\pm$ 0.21 & 1.63 $\pm$ 0.13 & 1.68 $\pm$ 0.20 & 10.43 $\pm$ 2.62 \\
1058-00205-1 & 188369 & +08 4263 & 1.20 $\pm$ 0.18 & 9.76 $\pm$ 0.21 & 1.48 $\pm$ 0.16 & 1.44 $\pm$ 0.29 & 7.79 $\pm$ 2.48 \\
1058-00635-1 & 188004 & +08 4249 & 1.31 $\pm$ 0.23 & 9.57 $\pm$ 0.24 & 1.69 $\pm$ 0.15 & - & 10.10 $\pm$ 2.25 \\
1058-00979-1 & - & +08 4223 & 1.19 $\pm$ 0.19 & 9.82 $\pm$ 0.22 & 1.38 $\pm$ 0.18 & - & 6.73 $\pm$ 2.04 \\
1058-01279-1 & 188237 & +08 4259 & 1.36 $\pm$ 0.17 & 9.51 $\pm$ 0.18 & 1.70 $\pm$ 0.10 & - & 10.45 $\pm$ 1.52 \\
1058-01537-1 & 187377 & +08 4228 & 1.35 $\pm$ 0.26 & 9.55 $\pm$ 0.25 & 1.61 $\pm$ 0.20 & - & 8.77 $\pm$ 2.43 \\
1058-01931-1 & - & - & 2.54 $\pm$ 0.10 & 8.80 $\pm$ 0.04 & 1.72 $\pm$ 0.06 & - & 9.20 $\pm$ 0.73 \\
1058-02389-1 & 188124 & +08 4252 & 1.37 $\pm$ 0.16 & 9.56 $\pm$ 0.14 & 1.58 $\pm$ 0.13 & - & 8.72 $\pm$ 1.85 \\
1058-02865-1 & 188214 & +08 4258 & 1.00 $\pm$ 0.10 & 9.99 $\pm$ 0.13 & 1.20 $\pm$ 0.10 & - & 5.45 $\pm$ 0.94 \\
1058-02919-1 & - & - & 1.37 $\pm$ 0.22 & 9.51 $\pm$ 0.21 & 1.46 $\pm$ 0.15 & - & 7.55 $\pm$ 1.54 \\
1058-02972-1 & 187094 & +07 4230 & 1.21 $\pm$ 0.22 & 9.72 $\pm$ 0.24 & 1.67 $\pm$ 0.13 & - & 10.38 $\pm$ 2.23 \\
1058-03032-1 & 187526 & +08 4230 & 1.85 $\pm$ 0.18 & 9.14 $\pm$ 0.10 & 1.36 $\pm$ 0.10 & - & 6.14 $\pm$ 0.86 \\
1058-03402-1 & - & +08 4232 & 1.23 $\pm$ 0.16 & 9.78 $\pm$ 0.20 & 1.43 $\pm$ 0.11 & - & 8.00 $\pm$ 1.39 \\
1062-00017-1 & 187552 & +09 4280 & 3.11 $\pm$ 0.62 & 8.58 $\pm$ 0.23 & 2.35 $\pm$ 0.35 & - & 24.51 $\pm$ 14.82 \\
1208-00063-1 & - & - & 1.02 $\pm$ 0.13 & 9.91 $\pm$ 0.18 & 1.58 $\pm$ 0.10 & - & 8.71 $\pm$ 1.43 \\
1208-00261-1 & 10364 & +19 277 & 1.37 $\pm$ 0.16 & 9.60 $\pm$ 0.17 & 1.48 $\pm$ 0.17 & 1.59 $\pm$ 0.22 & 8.91 $\pm$ 2.41 \\
1208-00623-1 & - & +18 219 & 1.13 $\pm$ 0.17 & 9.87 $\pm$ 0.21 & 2.14 $\pm$ 0.12 & - & 18.17 $\pm$ 3.96 \\
1211-00015-1 & - & - & 1.09 $\pm$ 0.10 & 9.78 $\pm$ 0.14 & 0.96 $\pm$ 0.09 & - & 3.91 $\pm$ 0.52 \\
1211-00016-1 & - & - & 0.98 $\pm$ 0.08 & 10.02 $\pm$ 0.11 & 1.27 $\pm$ 0.07 & - & 5.97 $\pm$ 0.80 \\
1211-00103-1 & - & - & 1.05 $\pm$ 0.14 & 9.93 $\pm$ 0.18 & 1.48 $\pm$ 0.08 & - & 8.18 $\pm$ 1.11 \\
1211-00153-1 & - & - & 1.83 $\pm$ 0.14 & 9.21 $\pm$ 0.07 & 1.67 $\pm$ 0.08 & - & 9.12 $\pm$ 1.27 \\
1211-00335-1 & - & +19 271 & 1.06 $\pm$ 0.15 & 9.90 $\pm$ 0.20 & 2.13 $\pm$ 0.10 & - & 17.43 $\pm$ 3.34 \\
1211-00406-1 & - & - & 1.75 $\pm$ 0.16 & 9.26 $\pm$ 0.11 & 1.65 $\pm$ 0.12 & - & 9.06 $\pm$ 1.50 \\
1211-00449-1 & - & - & 1.27 $\pm$ 0.09 & 9.65 $\pm$ 0.11 & 0.71 $\pm$ 0.07 & - & 3.01 $\pm$ 0.35 \\
1211-00603-1 & - & +20 274 & 1.02 $\pm$ 0.14 & 9.91 $\pm$ 0.19 & 2.24 $\pm$ 0.09 & - & 20.38 $\pm$ 3.48 \\
1211-00677-1 & - & - & 1.07 $\pm$ 0.15 & 9.89 $\pm$ 0.19 & 1.96 $\pm$ 0.10 & - & 15.36 $\pm$ 2.51 \\
1422-00100-1 & - & +20 2454 & 1.17 $\pm$ 0.14 & 9.89 $\pm$ 0.18 & 1.31 $\pm$ 0.12 & - & 6.87 $\pm$ 1.31 \\
1422-00614-1 & - & - & 1.16 $\pm$ 0.18 & 9.77 $\pm$ 0.22 & 1.36 $\pm$ 0.16 & - & 6.80 $\pm$ 1.66 \\
1422-00790-1 & - & +20 2457 & 0.96 $\pm$ 0.14 & 9.90 $\pm$ 0.19 & 2.81 $\pm$ 0.08 & - & 39.10 $\pm$ 6.23 \\
1422-01044-1 & - & - & 1.27 $\pm$ 0.33 & 9.70 $\pm$ 0.30 & 2.17 $\pm$ 0.18 & - & 18.94 $\pm$ 6.00 \\
1423-00013-1 & - & - & 1.26 $\pm$ 0.31 & 9.72 $\pm$ 0.29 & 2.08 $\pm$ 0.17 & - & 16.96 $\pm$ 5.32 \\
1423-00156-1 & - & - & 1.06 $\pm$ 0.13 & 9.95 $\pm$ 0.16 & 1.92 $\pm$ 0.09 & - & 14.53 $\pm$ 2.21 \\
1423-00165-1 & - & +20 2464 & 1.04 $\pm$ 0.07 & 10.03 $\pm$ 0.10 & 1.65 $\pm$ 0.06 & - & 10.06 $\pm$ 1.25 \\
1423-00248-1 & - & - & 1.22 $\pm$ 0.23 & 9.80 $\pm$ 0.25 & 1.76 $\pm$ 0.16 & - & 11.34 $\pm$ 3.15 \\
1423-00364-1 & - & - & 1.11 $\pm$ 0.18 & 9.84 $\pm$ 0.22 & 1.61 $\pm$ 0.13 & - & 9.07 $\pm$ 2.15 \\
1423-00457-1 & 89930 & +19 2335 & 1.54 $\pm$ 0.19 & 9.37 $\pm$ 0.14 & 1.47 $\pm$ 0.14 & 1.78 $\pm$ 0.18 & 9.89 $\pm$ 2.08 \\
1425-01506-1 & 89196 & +20 2460 & 1.33 $\pm$ 0.47 & 9.64 $\pm$ 0.35 & 2.42 $\pm$ 0.22 & - & 23.02 $\pm$ 9.73 \\
1426-00662-1 & - & +21 2173 & 1.05 $\pm$ 0.10 & 9.97 $\pm$ 0.14 & 1.47 $\pm$ 0.09 & - & 7.48 $\pm$ 1.32 \\
1426-00810-1 & 89772 & +20 2475 & 1.32 $\pm$ 0.36 & 9.62 $\pm$ 0.32 & 2.10 $\pm$ 0.18 & - & 18.14 $\pm$ 5.47 \\
1426-01004-1 & - & - & 1.99 $\pm$ 0.06 & 9.11 $\pm$ 0.05 & 1.58 $\pm$ 0.04 & - & 8.03 $\pm$ 0.66 \\
1426-01209-1 & 89471 & +21 2175 & 1.36 $\pm$ 0.30 & 9.63 $\pm$ 0.28 & 1.99 $\pm$ 0.16 & 2.39 $\pm$ 0.44 & 20.78 $\pm$ 10.07 \\
1496-00050-1 & - & - & 1.55 $\pm$ 0.40 & 9.27 $\pm$ 0.32 & 1.88 $\pm$ 0.19 & - & 11.00 $\pm$ 2.98 \\
1496-00290-1 & - & - & 1.07 $\pm$ 0.17 & 9.87 $\pm$ 0.21 & 1.69 $\pm$ 0.13 & - & 10.21 $\pm$ 2.28 \\
1496-00374-1 & - & +15 2932 & 1.40 $\pm$ 0.31 & 9.60 $\pm$ 0.28 & 1.71 $\pm$ 0.23 & - & 11.39 $\pm$ 5.99 \\
1496-00572-1 & 143257 & +16 2855 & 1.20 $\pm$ 0.17 & 9.82 $\pm$ 0.20 & 1.70 $\pm$ 0.09 & 2.19 $\pm$ 0.28 & 15.79 $\pm$ 4.95 \\
1496-00637-1 & 143064 & +16 2851 & 1.38 $\pm$ 0.21 & 9.64 $\pm$ 0.20 & 1.43 $\pm$ 0.17 & 1.89 $\pm$ 0.45 & 10.94 $\pm$ 5.07 \\
1496-00840-1 & - & +16 2852 & 1.34 $\pm$ 0.19 & 9.51 $\pm$ 0.20 & 1.72 $\pm$ 0.12 & - & 10.47 $\pm$ 1.90 \\
1496-00961-1 & - & - & 1.06 $\pm$ 0.11 & 9.88 $\pm$ 0.15 & 0.85 $\pm$ 0.11 & - & 3.53 $\pm$ 0.58 \\
1496-01002-1 & - & - & 0.97 $\pm$ 0.12 & 9.84 $\pm$ 0.42 & 0.04 $\pm$ 0.38 & - & 1.41 $\pm$ 0.51 \\
1496-01016-1 & - & +16 2854 & 1.19 $\pm$ 0.38 & 9.63 $\pm$ 0.36 & 1.66 $\pm$ 0.38 & - & 7.84 $\pm$ 4.82 \\
1496-01656-1 & - & - & 0.95 $\pm$ 0.08 & 10.00 $\pm$ 0.12 & 2.40 $\pm$ 0.05 & - & 26.89 $\pm$ 2.73 \\
1496-01841-1 & 142245 & +15 2925 & 1.42 $\pm$ 0.10 & 9.54 $\pm$ 0.08 & 0.59 $\pm$ 0.12 & 1.19 $\pm$ 0.15 & 4.18 $\pm$ 0.82 \\
1503-01050-1 & - & +15 2940 & 1.17 $\pm$ 0.18 & 9.74 $\pm$ 0.22 & 1.40 $\pm$ 0.14 & 1.88 $\pm$ 0.65 & 9.87 $\pm$ 5.71 \\
2818-00449-1 & - & +39 373 & 1.12 $\pm$ 0.12 & 9.92 $\pm$ 0.15 & 1.02 $\pm$ 0.11 & 1.28 $\pm$ 0.41 & 5.57 $\pm$ 2.12 \\
2818-00504-1 & - & +40 334 & 1.05 $\pm$ 0.18 & 9.84 $\pm$ 0.23 & 2.50 $\pm$ 0.12 & 2.33 $\pm$ 0.52 & 26.34 $\pm$ 11.70 \\
2818-00602-1 & - & - & 3.50 $\pm$ 0.81 & 8.39 $\pm$ 0.27 & 2.87 $\pm$ 0.29 & - & 47.65 $\pm$ 27.90 \\
2818-00733-1 & - & +40 317 & 1.17 $\pm$ 0.18 & 9.82 $\pm$ 0.21 & 1.57 $\pm$ 0.10 & 2.04 $\pm$ 0.37 & 13.38 $\pm$ 4.79 \\
2818-00874-1 & - & +40 338 & 0.98 $\pm$ 0.08 & 10.00 $\pm$ 0.12 & 1.36 $\pm$ 0.07 & - & 6.88 $\pm$ 0.81 \\
2818-00990-1 & - & - & 3.56 $\pm$ 0.54 & 8.40 $\pm$ 0.18 & 2.68 $\pm$ 0.31 & - & 41.00 $\pm$ 24.01 \\
2818-01153-1 & - & +40 321 & 1.13 $\pm$ 0.21 & 9.79 $\pm$ 0.25 & 2.39 $\pm$ 0.13 & - & 24.59 $\pm$ 5.83 \\
2818-02188-1 & 9712 & +40 328 & 1.60 $\pm$ 0.20 & 9.35 $\pm$ 0.14 & 1.18 $\pm$ 0.14 & 1.78 $\pm$ 0.12 & 8.84 $\pm$ 1.76 \\
2822-00208-1 & - & - & 1.42 $\pm$ 0.19 & 9.54 $\pm$ 0.18 & 1.51 $\pm$ 0.14 & - & 8.57 $\pm$ 1.63 \\
2822-00410-1 & 9416 & +41 300 & 1.87 $\pm$ 0.22 & 9.16 $\pm$ 0.12 & 1.78 $\pm$ 0.06 & 2.29 $\pm$ 0.66 & 14.70 $\pm$ 8.02 \\
2822-01573-1 & - & +41 306 & 2.33 $\pm$ 0.19 & 8.89 $\pm$ 0.11 & 1.86 $\pm$ 0.09 & - & 11.03 $\pm$ 1.42 \\
2822-01643-1 & 9519 & +41 304 & 1.13 $\pm$ 0.19 & 9.85 $\pm$ 0.22 & 1.98 $\pm$ 0.13 & - & 14.61 $\pm$ 3.41 \\
2822-02010-1 & - & - & 1.24 $\pm$ 0.09 & 9.75 $\pm$ 0.11 & 0.82 $\pm$ 0.08 & - & 3.65 $\pm$ 0.44 \\
2823-01028-1 & - & +40 351 & 0.88 $\pm$ 0.04 & 10.07 $\pm$ 0.06 & 1.88 $\pm$ 0.06 & 1.17 $\pm$ 0.34 & 7.67 $\pm$ 1.97 \\
2823-01398-1 & - & +40 348 & 0.96 $\pm$ 0.09 & 9.99 $\pm$ 0.13 & 1.36 $\pm$ 0.08 & - & 6.51 $\pm$ 0.94 \\
2823-01786-1 & 10455 & +40 352 & 1.79 $\pm$ 0.24 & 9.26 $\pm$ 0.16 & 1.58 $\pm$ 0.16 & - & 8.58 $\pm$ 1.72 \\
3011-00547-1 & - & +44 2037 & 1.04 $\pm$ 0.08 & 10.00 $\pm$ 0.11 & 0.88 $\pm$ 0.10 & - & 3.93 $\pm$ 0.58 \\
3011-00791-1 & 95127 & +44 2038 & 1.74 $\pm$ 0.37 & 9.35 $\pm$ 0.27 & 2.42 $\pm$ 0.24 & 2.13 $\pm$ 0.37 & 9.71 $\pm$ 4.44 \\
3012-00126-1 & - & +44 2059 & 1.29 $\pm$ 0.21 & 9.53 $\pm$ 0.22 & 1.72 $\pm$ 0.14 & - & 10.07 $\pm$ 1.97 \\
3012-00145-1 & 96992 & +44 2063 & 0.99 $\pm$ 0.11 & 9.96 $\pm$ 0.16 & 1.47 $\pm$ 0.09 & 1.88 $\pm$ 0.46 & 9.92 $\pm$ 4.19 \\
3012-00273-1 & - & +44 2057 & 0.88 $\pm$ 0.03 & 10.09 $\pm$ 0.05 & 0.42 $\pm$ 0.04 & - & 1.74 $\pm$ 0.18 \\
3012-00285-1 & - & +44 2047 & 1.17 $\pm$ 0.17 & 9.89 $\pm$ 0.20 & 1.47 $\pm$ 0.14 & - & 7.91 $\pm$ 2.15 \\
3012-00470-1 & - & - & 1.07 $\pm$ 0.12 & 9.88 $\pm$ 0.17 & 0.96 $\pm$ 0.13 & - & 4.09 $\pm$ 0.81 \\
3012-00667-1 & 96127 & +45 1892 & 1.39 $\pm$ 0.29 & 9.54 $\pm$ 0.27 & 2.33 $\pm$ 0.15 & 2.76 $\pm$ 0.34 & 32.26 $\pm$ 13.62 \\
3012-00676-1 & - & +44 2041 & 1.11 $\pm$ 0.15 & 9.88 $\pm$ 0.19 & 1.24 $\pm$ 0.14 & - & 5.96 $\pm$ 1.34 \\
3012-01263-1 & - & - & 1.16 $\pm$ 0.20 & 9.76 $\pm$ 0.24 & 2.17 $\pm$ 0.12 & - & 20.71 $\pm$ 3.85 \\
3012-01504-1 & - & +43 2070 & 1.00 $\pm$ 0.05 & 10.05 $\pm$ 0.08 & 0.90 $\pm$ 0.11 & 0.64 $\pm$ 0.22 & 3.37 $\pm$ 0.78 \\
3012-01520-1 & - & +43 2080 & 1.31 $\pm$ 0.39 & 9.67 $\pm$ 0.32 & 2.25 $\pm$ 0.19 & - & 21.09 $\pm$ 7.44 \\
3012-02518-1 & 95296 & +43 2069 & 2.03 $\pm$ 0.33 & 9.11 $\pm$ 0.20 & 1.66 $\pm$ 0.16 & 2.11 $\pm$ 0.16 & 15.02 $\pm$ 3.84 \\
3012-02527-1 & - & +44 2046 & 1.10 $\pm$ 0.11 & 9.92 $\pm$ 0.15 & 0.69 $\pm$ 0.12 & - & 3.10 $\pm$ 0.59 \\
3018-00288-1 & - & +41 2310 & 0.98 $\pm$ 0.09 & 9.99 $\pm$ 0.13 & 1.61 $\pm$ 0.08 & - & 8.90 $\pm$ 1.37 \\
3018-00336-1 & - & +41 2304 & 1.18 $\pm$ 0.18 & 9.86 $\pm$ 0.21 & 1.48 $\pm$ 0.14 & - & 7.83 $\pm$ 2.12 \\
3018-00350-1 & 108872 & +41 2298 & 1.23 $\pm$ 0.17 & 9.74 $\pm$ 0.20 & 1.57 $\pm$ 0.15 & 1.31 $\pm$ 0.31 & 7.78 $\pm$ 2.50 \\
3018-00704-1 & 109461 & +41 2305 & 1.51 $\pm$ 0.36 & 9.36 $\pm$ 0.28 & 1.89 $\pm$ 0.20 & 2.03 $\pm$ 0.56 & 13.79 $\pm$ 6.96 \\
3018-00708-1 & 110065 & +41 2313 & 1.44 $\pm$ 0.11 & 9.53 $\pm$ 0.08 & 0.60 $\pm$ 0.13 & 0.88 $\pm$ 0.17 & 3.25 $\pm$ 0.68 \\
3018-00996-1 & 109681 & +41 2308 & 1.27 $\pm$ 0.13 & 9.73 $\pm$ 0.15 & 0.95 $\pm$ 0.15 & 1.28 $\pm$ 0.16 & 5.41 $\pm$ 1.30 \\
3018-01050-1 & 109740 & +41 2309 & 1.12 $\pm$ 0.17 & 9.86 $\pm$ 0.20 & 1.55 $\pm$ 0.11 & 1.68 $\pm$ 0.43 & 9.46 $\pm$ 3.97 \\
3020-01183-1 & - & +42 2322 & 1.07 $\pm$ 0.09 & 9.99 $\pm$ 0.12 & 1.61 $\pm$ 0.13 & 0.85 $\pm$ 0.36 & 7.10 $\pm$ 1.83 \\
3020-01288-1 & - & +42 2315 & 1.38 $\pm$ 0.30 & 9.56 $\pm$ 0.29 & 1.99 $\pm$ 0.26 & - & 17.85 $\pm$ 10.01 \\
3020-02438-1 & - & +42 2320 & 1.28 $\pm$ 0.17 & 9.61 $\pm$ 0.18 & 1.48 $\pm$ 0.11 & - & 8.00 $\pm$ 1.21 \\
3105-00228-1 & - & +39 3457 & 1.14 $\pm$ 0.21 & 9.71 $\pm$ 0.26 & 1.77 $\pm$ 0.15 & - & 10.60 $\pm$ 2.55 \\
3105-00535-1 & - & +38 3228 & 1.31 $\pm$ 0.18 & 9.67 $\pm$ 0.19 & 1.51 $\pm$ 0.12 & - & 8.92 $\pm$ 1.56 \\
3105-00683-1 & - & +38 3201 & 0.98 $\pm$ 0.08 & 10.02 $\pm$ 0.11 & 1.12 $\pm$ 0.08 & - & 5.01 $\pm$ 0.71 \\
3105-00692-1 & - & - & 1.21 $\pm$ 0.29 & 9.65 $\pm$ 0.29 & 2.19 $\pm$ 0.14 & - & 20.08 $\pm$ 4.25 \\
3105-00873-1 & - & - & 1.13 $\pm$ 0.19 & 9.82 $\pm$ 0.23 & 2.00 $\pm$ 0.11 & - & 16.67 $\pm$ 2.98 \\
3105-01095-1 & - & +37 3182 & 1.93 $\pm$ 0.21 & 9.17 $\pm$ 0.11 & 1.73 $\pm$ 0.08 & - & 9.97 $\pm$ 1.45 \\
3105-01103-1 & - & +37 3172 & 3.75 $\pm$ 0.86 & 8.38 $\pm$ 0.28 & 2.84 $\pm$ 0.28 & - & 50.24 $\pm$ 28.97 \\
3105-01137-1 & - & +38 3205 & 0.97 $\pm$ 0.08 & 10.02 $\pm$ 0.11 & 1.50 $\pm$ 0.09 & - & 6.89 $\pm$ 1.16 \\
3105-01851-1 & - & +37 3198 & 1.45 $\pm$ 0.35 & 9.48 $\pm$ 0.30 & 1.81 $\pm$ 0.16 & - & 11.96 $\pm$ 2.96 \\
3105-01862-1 & - & - & 1.48 $\pm$ 0.23 & 9.45 $\pm$ 0.17 & 1.80 $\pm$ 0.06 & - & 11.28 $\pm$ 1.47 \\
3105-02077-1 & - & +38 3235 & 0.94 $\pm$ 0.03 & 10.08 $\pm$ 0.05 & 1.25 $\pm$ 0.04 & - & 5.74 $\pm$ 0.50 \\
3109-00661-1 & - & +39 3464 & 1.13 $\pm$ 0.12 & 9.86 $\pm$ 0.15 & 0.98 $\pm$ 0.09 & - & 4.42 $\pm$ 0.60 \\
3109-01946-1 & - & - & 0.99 $\pm$ 0.12 & 9.95 $\pm$ 0.16 & 1.46 $\pm$ 0.10 & - & 7.41 $\pm$ 1.19 \\
3109-02342-1 & - & +39 3480 & 1.09 $\pm$ 0.17 & 9.83 $\pm$ 0.22 & 1.55 $\pm$ 0.12 & - & 8.49 $\pm$ 1.56 \\
3118-00440-1 & - & +39 3493 & 2.02 $\pm$ 0.32 & 9.10 $\pm$ 0.17 & 1.76 $\pm$ 0.12 & - & 10.06 $\pm$ 2.05 \\
3118-02068-1 & - & +38 3258 & 1.16 $\pm$ 0.16 & 9.89 $\pm$ 0.19 & 1.48 $\pm$ 0.11 & - & 8.88 $\pm$ 1.70 \\
3226-00556-1 & 215443 & +43 4293 & 1.94 $\pm$ 0.28 & 9.12 $\pm$ 0.15 & 1.82 $\pm$ 0.10 & - & 10.86 $\pm$ 1.83 \\
3226-00696-1 & - & +43 4303 & 1.16 $\pm$ 0.18 & 9.85 $\pm$ 0.22 & 1.70 $\pm$ 0.12 & - & 11.10 $\pm$ 2.21 \\
3226-00868-1 & 215040 & +43 4270 & 1.28 $\pm$ 0.43 & 9.64 $\pm$ 0.38 & 1.49 $\pm$ 0.70 & - & 8.11 $\pm$ 6.94 \\
3226-00993-1 & 215472 & +43 4295 & 1.57 $\pm$ 0.22 & 9.40 $\pm$ 0.18 & 1.52 $\pm$ 0.13 & - & 8.37 $\pm$ 1.40 \\
3226-00997-1 & 215335 & +43 4288 & 1.30 $\pm$ 0.18 & 9.67 $\pm$ 0.19 & 1.55 $\pm$ 0.10 & 1.81 $\pm$ 0.18 & 11.45 $\pm$ 2.45 \\
3226-01219-1 & 216016 & +43 4310 & 1.54 $\pm$ 0.23 & 9.35 $\pm$ 0.18 & 1.77 $\pm$ 0.10 & - & 10.31 $\pm$ 1.59 \\
3226-01373-1 & 215576 & +43 4299 & 1.28 $\pm$ 0.23 & 9.69 $\pm$ 0.24 & 1.35 $\pm$ 0.19 & - & 6.98 $\pm$ 1.86 \\
3226-01589-1 & - & - & 1.08 $\pm$ 0.13 & 9.88 $\pm$ 0.17 & 1.21 $\pm$ 0.09 & - & 5.78 $\pm$ 0.82 \\
3226-01858-1 & 216161 & +42 4506 & 1.14 $\pm$ 0.13 & 9.88 $\pm$ 0.17 & 1.13 $\pm$ 0.11 & - & 5.46 $\pm$ 0.90 \\
3226-02051-1 & 215346 & +43 4289 & 1.08 $\pm$ 0.12 & 9.95 $\pm$ 0.16 & 1.88 $\pm$ 0.08 & - & 14.10 $\pm$ 2.05 \\
3226-02100-1 & 215897 & +42 4501 & 1.40 $\pm$ 0.20 & 9.53 $\pm$ 0.18 & 1.28 $\pm$ 0.11 & - & 6.29 $\pm$ 0.97 \\
3226-02285-1 & 214868 & +43 4266 & 1.38 $\pm$ 0.17 & 9.51 $\pm$ 0.17 & 2.31 $\pm$ 0.14 & 2.42 $\pm$ 0.10 & 27.70 $\pm$ 5.00 \\
3227-00068-1 & 216307 & +42 4517 & 1.66 $\pm$ 0.29 & 9.34 $\pm$ 0.20 & 1.73 $\pm$ 0.12 & - & 10.20 $\pm$ 2.12 \\
3227-00213-1 & 216257 & +44 4251 & 1.01 $\pm$ 0.10 & 9.98 $\pm$ 0.14 & 2.01 $\pm$ 0.07 & 2.06 $\pm$ 0.32 & 16.41 $\pm$ 4.95 \\
3300-00133-1 & - & - & 2.04 $\pm$ 0.16 & 9.08 $\pm$ 0.10 & 1.41 $\pm$ 0.10 & - & 6.76 $\pm$ 0.92 \\
3300-01380-1 & - & +47 678 & 2.05 $\pm$ 0.15 & 9.08 $\pm$ 0.09 & 1.62 $\pm$ 0.08 & - & 8.09 $\pm$ 1.26 \\
3300-01645-1 & - & +47 682 & 1.25 $\pm$ 0.28 & 9.72 $\pm$ 0.28 & 2.00 $\pm$ 0.16 & - & 16.37 $\pm$ 4.19 \\
3300-01952-1 & - & +48 749 & 1.45 $\pm$ 0.27 & 9.52 $\pm$ 0.24 & 1.57 $\pm$ 0.21 & - & 8.73 $\pm$ 2.43 \\
3304-00088-1 & - & - & 1.66 $\pm$ 0.26 & 9.28 $\pm$ 0.18 & 1.77 $\pm$ 0.12 & - & 10.09 $\pm$ 2.03 \\
3304-00090-1 & - & +48 740 & 1.12 $\pm$ 0.18 & 9.87 $\pm$ 0.21 & 1.64 $\pm$ 0.12 & 2.49 $\pm$ 0.84 & 19.27 $\pm$ 15.02 \\
3304-00101-1 & 17092 & +49 767 & 1.25 $\pm$ 0.17 & 9.73 $\pm$ 0.19 & 1.57 $\pm$ 0.09 & - & 9.68 $\pm$ 1.40 \\
3304-00110-1 & - & +48 725 & 1.21 $\pm$ 0.22 & 9.80 $\pm$ 0.24 & 2.08 $\pm$ 0.13 & - & 18.48 $\pm$ 4.16 \\
3304-00323-1 & - & +48 738 & 1.16 $\pm$ 0.22 & 9.79 $\pm$ 0.25 & 1.97 $\pm$ 0.14 & - & 14.97 $\pm$ 3.50 \\
3304-00405-1 & - & +49 775 & 1.47 $\pm$ 0.48 & 9.60 $\pm$ 0.36 & 2.06 $\pm$ 0.23 & - & 17.26 $\pm$ 6.97 \\
3304-00408-1 & 17028 & +48 750 & 1.35 $\pm$ 0.41 & 9.59 $\pm$ 0.32 & 2.53 $\pm$ 0.24 & 2.58 $\pm$ 0.59 & 30.00 $\pm$ 20.83 \\
3304-00479-1 & 16992 & +49 758 & 3.62 $\pm$ 0.79 & 8.39 $\pm$ 0.26 & 2.77 $\pm$ 0.29 & - & 45.95 $\pm$ 27.10 \\
3304-00553-1 & - & - & 1.31 $\pm$ 0.34 & 9.70 $\pm$ 0.30 & 2.08 $\pm$ 0.18 & - & 17.28 $\pm$ 5.61 \\
3304-01910-1 & - & +49 772 & 1.23 $\pm$ 0.27 & 9.73 $\pm$ 0.28 & 1.90 $\pm$ 0.16 & - & 14.12 $\pm$ 3.59 \\
3314-01371-1 & - & - & 1.30 $\pm$ 0.16 & 9.59 $\pm$ 0.15 & 1.02 $\pm$ 0.11 & - & 4.39 $\pm$ 0.75 \\
3318-00789-1 & - & +49 835 & 1.00 $\pm$ 0.12 & 9.94 $\pm$ 0.17 & 1.44 $\pm$ 0.11 & - & 7.02 $\pm$ 1.31 \\
3318-01302-1 & - & +49 852 & 1.95 $\pm$ 0.69 & 9.16 $\pm$ 0.40 & 2.25 $\pm$ 0.26 & - & 21.93 $\pm$ 8.10 \\
3318-01333-1 & - & - & 1.19 $\pm$ 0.14 & 9.78 $\pm$ 0.18 & 1.21 $\pm$ 0.11 & - & 5.90 $\pm$ 1.00 \\
3318-01427-1 & - & - & 1.54 $\pm$ 0.21 & 9.41 $\pm$ 0.16 & 1.68 $\pm$ 0.11 & - & 9.96 $\pm$ 1.54 \\
3318-01487-1 & - & +49 859 & 1.72 $\pm$ 0.24 & 9.29 $\pm$ 0.17 & 1.63 $\pm$ 0.16 & - & 8.87 $\pm$ 1.76 \\
3318-01515-1 & - & +49 828 & 1.52 $\pm$ 0.22 & 9.37 $\pm$ 0.18 & 1.47 $\pm$ 0.13 & - & 7.55 $\pm$ 1.28 \\
3318-01538-1 & 18927 & +49 838 & 1.98 $\pm$ 0.34 & 9.13 $\pm$ 0.19 & 1.74 $\pm$ 0.14 & - & 9.97 $\pm$ 2.16 \\
3319-00170-1 & - & - & 1.09 $\pm$ 0.15 & 9.86 $\pm$ 0.20 & 1.46 $\pm$ 0.12 & - & 7.81 $\pm$ 1.38 \\
3319-00172-1 & - & +49 872 & 1.57 $\pm$ 0.82 & 9.44 $\pm$ 0.48 & 2.49 $\pm$ 0.27 & - & 30.62 $\pm$ 13.98 \\
3319-00366-1 & 19636 & +48 859 & 1.36 $\pm$ 0.24 & 9.54 $\pm$ 0.23 & 1.59 $\pm$ 0.18 & - & 8.82 $\pm$ 2.14 \\
3430-00053-1 & - & - & 0.97 $\pm$ 0.09 & 9.98 $\pm$ 0.13 & 1.18 $\pm$ 0.09 & - & 5.17 $\pm$ 0.79 \\
3430-00480-1 & - & +52 1375 & 1.00 $\pm$ 0.12 & 9.93 $\pm$ 0.16 & 1.67 $\pm$ 0.09 & - & 9.61 $\pm$ 1.64 \\
3430-00683-1 & - & - & 0.94 $\pm$ 0.08 & 10.01 $\pm$ 0.12 & 1.58 $\pm$ 0.09 & - & 8.27 $\pm$ 1.20 \\
3430-00747-1 & 233601 & +53 1310 & 1.13 $\pm$ 0.21 & 9.80 $\pm$ 0.25 & 2.44 $\pm$ 0.13 & - & 25.30 $\pm$ 6.12 \\
3431-00086-1 & - & - & 3.54 $\pm$ 0.82 & 8.39 $\pm$ 0.27 & 2.86 $\pm$ 0.30 & - & 47.62 $\pm$ 28.17 \\
3431-00680-1 & - & - & 1.36 $\pm$ 0.26 & 9.54 $\pm$ 0.25 & 1.72 $\pm$ 0.14 & - & 10.37 $\pm$ 2.30 \\
3621-00326-1 & 215909 & +44 4234 & 1.22 $\pm$ 0.21 & 9.76 $\pm$ 0.23 & 2.03 $\pm$ 0.16 & 1.96 $\pm$ 0.32 & 16.35 $\pm$ 5.61 \\
3621-00445-1 & - & +44 4237 & 1.26 $\pm$ 0.18 & 9.76 $\pm$ 0.20 & 1.50 $\pm$ 0.12 & - & 8.88 $\pm$ 1.79 \\
3621-00792-1 & 215150 & +44 4203 & 1.12 $\pm$ 0.17 & 9.86 $\pm$ 0.20 & 1.17 $\pm$ 0.18 & - & 5.30 $\pm$ 1.54 \\
3663-00024-1 & - & - & 1.11 $\pm$ 0.14 & 9.91 $\pm$ 0.18 & 1.54 $\pm$ 0.09 & - & 9.42 $\pm$ 1.52 \\
3663-00578-1 & - & - & 1.47 $\pm$ 0.13 & 9.33 $\pm$ 0.11 & 1.83 $\pm$ 0.07 & - & 11.44 $\pm$ 1.12 \\
3663-00622-1 & 236545 & +56 135 & 1.80 $\pm$ 0.26 & 9.23 $\pm$ 0.15 & 1.77 $\pm$ 0.07 & 2.33 $\pm$ 0.84 & 15.44 $\pm$ 10.87 \\
3663-00654-1 & - & - & 1.03 $\pm$ 0.12 & 9.93 $\pm$ 0.16 & 1.59 $\pm$ 0.08 & - & 9.15 $\pm$ 1.38 \\
3663-00789-1 & 236525 & +56 127 & 1.95 $\pm$ 0.22 & 9.17 $\pm$ 0.12 & 1.67 $\pm$ 0.10 & - & 9.26 $\pm$ 1.50 \\
3663-00838-1 & 236555 & +56 138 & 3.01 $\pm$ 0.63 & 8.59 $\pm$ 0.24 & 2.39 $\pm$ 0.36 & - & 24.78 $\pm$ 15.11 \\
3663-01007-1 & 236559 & +56 140 & 1.05 $\pm$ 0.10 & 9.98 $\pm$ 0.14 & 1.42 $\pm$ 0.09 & - & 7.52 $\pm$ 1.10 \\
3663-01040-1 & 236565 & +56 142 & 1.34 $\pm$ 0.18 & 9.56 $\pm$ 0.19 & 1.54 $\pm$ 0.13 & 2.28 $\pm$ 0.60 & 14.62 $\pm$ 8.08 \\
3663-01463-1 & - & +55 182 & 0.92 $\pm$ 0.08 & 10.00 $\pm$ 0.13 & 1.64 $\pm$ 0.08 & - & 8.32 $\pm$ 1.29 \\
3663-01888-1 & - & - & 1.05 $\pm$ 0.13 & 9.90 $\pm$ 0.18 & 1.85 $\pm$ 0.09 & - & 13.32 $\pm$ 1.80 \\
3663-01966-1 & - & - & 2.88 $\pm$ 0.21 & 8.64 $\pm$ 0.11 & 2.26 $\pm$ 0.09 & - & 17.66 $\pm$ 2.28 \\
3663-01992-1 & - & - & 1.20 $\pm$ 0.10 & 9.69 $\pm$ 0.12 & 0.85 $\pm$ 0.10 & - & 3.51 $\pm$ 0.55 \\
3663-02054-1 & - & - & 1.09 $\pm$ 0.15 & 9.86 $\pm$ 0.20 & 1.52 $\pm$ 0.12 & - & 8.22 $\pm$ 1.62 \\
3663-02059-1 & - & +56 139A & 1.62 $\pm$ 0.20 & 9.37 $\pm$ 0.14 & 1.74 $\pm$ 0.08 & - & 10.51 $\pm$ 1.56 \\
3663-02434-1 & 236543 & +56 134 & 1.17 $\pm$ 0.17 & 9.74 $\pm$ 0.22 & 2.06 $\pm$ 0.30 & 1.83 $\pm$ 0.50 & 13.21 $\pm$ 5.85 \\
3667-00262-1 & 236563 & +57 163 & 1.61 $\pm$ 0.26 & 9.31 $\pm$ 0.18 & 1.83 $\pm$ 0.06 & - & 11.06 $\pm$ 1.47 \\
3667-00512-1 & 236530 & +57 146 & 1.10 $\pm$ 0.17 & 9.86 $\pm$ 0.20 & 1.53 $\pm$ 0.12 & - & 8.19 $\pm$ 1.77 \\
3667-00550-1 & - & - & 1.79 $\pm$ 0.18 & 9.18 $\pm$ 0.13 & 1.55 $\pm$ 0.11 & - & 7.70 $\pm$ 1.12 \\
3667-01178-1 & 3933 & +57 130 & 2.05 $\pm$ 0.28 & 9.07 $\pm$ 0.15 & 1.85 $\pm$ 0.11 & 1.86 $\pm$ 0.35 & 11.28 $\pm$ 3.50 \\
3667-01280-1 & - & - & 1.87 $\pm$ 0.17 & 9.13 $\pm$ 0.10 & 1.38 $\pm$ 0.10 & - & 6.26 $\pm$ 0.86 \\
3667-01442-1 & - & - & 1.92 $\pm$ 0.16 & 9.17 $\pm$ 0.09 & 1.70 $\pm$ 0.07 & - & 9.56 $\pm$ 1.11 \\
3667-01636-1 & - & +57 144 & 1.40 $\pm$ 0.23 & 9.49 $\pm$ 0.19 & 1.02 $\pm$ 0.15 & 2.33 $\pm$ 0.90 & 11.79 $\pm$ 10.57 \\
3667-01656-1 & 236562 & +57 162 & 1.41 $\pm$ 0.23 & 9.55 $\pm$ 0.22 & 1.40 $\pm$ 0.19 & - & 7.36 $\pm$ 1.83 \\
3676-02387-1 & - & - & 1.35 $\pm$ 0.22 & 9.57 $\pm$ 0.20 & 0.77 $\pm$ 0.20 & - & 3.19 $\pm$ 0.94 \\
3805-00193-1 & 233604 & +54 1280 & 1.24 $\pm$ 0.14 & 9.62 $\pm$ 0.16 & 1.66 $\pm$ 0.09 & - & 9.81 $\pm$ 1.14 \\
3805-01162-1 & - & - & 1.07 $\pm$ 0.09 & 9.99 $\pm$ 0.13 & 0.95 $\pm$ 0.12 & - & 4.18 $\pm$ 0.86 \\
3806-00244-1 & - & - & 1.03 $\pm$ 0.12 & 9.91 $\pm$ 0.16 & 1.37 $\pm$ 0.09 & - & 6.95 $\pm$ 0.96 \\
3806-00861-1 & - & - & 1.00 $\pm$ 0.11 & 9.92 $\pm$ 0.16 & 1.61 $\pm$ 0.09 & - & 9.30 $\pm$ 1.24 \\
3806-01026-1 & - & - & 0.98 $\pm$ 0.03 & 10.08 $\pm$ 0.05 & 0.87 $\pm$ 0.05 & - & 3.83 $\pm$ 0.36 \\
3806-01071-1 & 233612 & +53 1318 & 1.07 $\pm$ 0.20 & 9.79 $\pm$ 0.25 & 2.77 $\pm$ 0.13 & - & 37.95 $\pm$ 8.83 \\
3806-01289-1 & 233615 & +53 1324 & 1.09 $\pm$ 0.18 & 9.85 $\pm$ 0.23 & 2.35 $\pm$ 0.12 & - & 21.43 $\pm$ 4.96 \\
3917-01107-1 & 238914 & +59 1909 & 1.47 $\pm$ 0.47 & 9.50 $\pm$ 0.38 & 1.85 $\pm$ 0.19 & - & 12.73 $\pm$ 3.89 \\
3917-01228-1 & - & - & 1.76 $\pm$ 0.21 & 9.29 $\pm$ 0.13 & 1.73 $\pm$ 0.07 & - & 10.42 $\pm$ 1.50 \\
3930-00143-1 & 238928 & +59 1916 & 1.12 $\pm$ 0.14 & 9.93 $\pm$ 0.18 & 1.70 $\pm$ 0.10 & - & 11.06 $\pm$ 1.91 \\
3930-00383-1 & - & - & 1.00 $\pm$ 0.12 & 9.92 $\pm$ 0.16 & 1.74 $\pm$ 0.09 & - & 11.19 $\pm$ 1.56 \\
3930-00519-1 & - & +59 1920 & 1.03 $\pm$ 0.13 & 9.93 $\pm$ 0.18 & 2.16 $\pm$ 0.09 & - & 18.44 $\pm$ 3.11 \\
3930-00524-1 & 174259 & +59 1921 & 1.44 $\pm$ 0.22 & 9.41 $\pm$ 0.21 & 1.87 $\pm$ 0.13 & 1.78 $\pm$ 0.14 & 11.20 $\pm$ 1.95 \\
3930-00551-1 & - & - & 1.10 $\pm$ 0.12 & 9.92 $\pm$ 0.16 & 1.53 $\pm$ 0.09 & - & 9.05 $\pm$ 1.37 \\
3930-00665-1 & - & - & 1.64 $\pm$ 0.15 & 9.26 $\pm$ 0.11 & 1.64 $\pm$ 0.08 & - & 9.18 $\pm$ 1.00 \\
3930-00681-1 & - & - & 1.25 $\pm$ 0.26 & 9.70 $\pm$ 0.26 & 1.38 $\pm$ 0.23 & - & 6.90 $\pm$ 2.29 \\
3930-00783-1 & 174062 & +59 1918 & 1.01 $\pm$ 0.11 & 9.98 $\pm$ 0.14 & 1.00 $\pm$ 0.09 & - & 4.37 $\pm$ 0.68 \\
3930-00952-1 & - & - & 1.25 $\pm$ 0.29 & 9.73 $\pm$ 0.28 & 2.02 $\pm$ 0.17 & - & 16.04 $\pm$ 4.70 \\
3930-01302-1 & - & - & 1.00 $\pm$ 0.08 & 9.99 $\pm$ 0.12 & 1.50 $\pm$ 0.07 & - & 8.35 $\pm$ 0.97 \\
3930-01530-1 & - & - & 1.03 $\pm$ 0.10 & 9.89 $\pm$ 0.15 & 1.56 $\pm$ 0.08 & - & 8.84 $\pm$ 0.97 \\
3930-01669-1 & - & +58 1834 & 1.02 $\pm$ 0.13 & 9.85 $\pm$ 0.19 & 1.74 $\pm$ 0.10 & - & 10.34 $\pm$ 1.70 \\
3993-00227-1 & 240189A & +56 2947A & 0.99 $\pm$ 0.10 & 9.93 $\pm$ 0.15 & 0.88 $\pm$ 0.12 & - & 3.54 $\pm$ 0.68 \\
3993-01107-1 & 240188 & +55 2896 & 2.06 $\pm$ 0.13 & 9.09 $\pm$ 0.08 & 1.60 $\pm$ 0.07 & - & 8.01 $\pm$ 1.16 \\
3993-01850-1 & - & - & 1.36 $\pm$ 0.08 & 9.56 $\pm$ 0.08 & 0.78 $\pm$ 0.10 & - & 2.67 $\pm$ 0.35 \\
4006-00019-1 & 240237 & +57 2714 & 1.46 $\pm$ 0.32 & 9.48 $\pm$ 0.29 & 2.37 $\pm$ 0.14 & - & 28.24 $\pm$ 6.54 \\
4006-00340-1 & - & - & 1.65 $\pm$ 0.24 & 9.32 $\pm$ 0.16 & 1.76 $\pm$ 0.08 & - & 10.58 $\pm$ 1.47 \\
4006-00629-1 & 219415 & +55 2926 & 1.04 $\pm$ 0.09 & 10.00 $\pm$ 0.12 & 0.72 $\pm$ 0.13 & - & 3.12 $\pm$ 0.65 \\
4006-00797-1 & 240226 & +55 2918 & 1.80 $\pm$ 0.15 & 9.20 $\pm$ 0.07 & 1.79 $\pm$ 0.04 & - & 10.88 $\pm$ 1.34 \\
4006-00832-1 & - & +56 2967 & 2.08 $\pm$ 0.16 & 9.05 $\pm$ 0.09 & 1.62 $\pm$ 0.08 & - & 8.11 $\pm$ 0.99 \\
4006-00890-1 & - & +56 2957 & 1.44 $\pm$ 0.31 & 9.60 $\pm$ 0.28 & 2.17 $\pm$ 0.25 & - & 22.55 $\pm$ 12.01 \\
4006-00980-1 & 240210 & +56 2959 & 1.30 $\pm$ 0.38 & 9.68 $\pm$ 0.32 & 2.14 $\pm$ 0.20 & - & 17.13 $\pm$ 6.26 \\
4006-01008-1 & 240224 & +56 2963 & 1.30 $\pm$ 0.29 & 9.64 $\pm$ 0.26 & 1.78 $\pm$ 0.08 & - & 11.10 $\pm$ 1.98 \\
4006-01039-1 & 219812 & +55 2937 & 2.02 $\pm$ 0.16 & 9.07 $\pm$ 0.08 & 1.63 $\pm$ 0.07 & - & 7.86 $\pm$ 1.20 \\
4006-01055-1 & - & - & 1.13 $\pm$ 0.18 & 9.78 $\pm$ 0.23 & 1.60 $\pm$ 0.13 & - & 9.08 $\pm$ 1.87 \\
4211-00383-1 & - & +67 1028 & 1.17 $\pm$ 0.19 & 9.82 $\pm$ 0.22 & 1.38 $\pm$ 0.18 & - & 6.75 $\pm$ 1.93 \\
4211-00438-1 & - & +67 1024 & 1.27 $\pm$ 0.22 & 9.69 $\pm$ 0.23 & 1.56 $\pm$ 0.15 & - & 8.95 $\pm$ 2.17 \\
4215-01352-1 & - & +60 1847 & 1.32 $\pm$ 0.38 & 9.67 $\pm$ 0.31 & 2.30 $\pm$ 0.18 & - & 23.62 $\pm$ 7.79 \\
4215-02018-1 & - & - & 1.54 $\pm$ 0.20 & 9.36 $\pm$ 0.16 & 1.54 $\pm$ 0.13 & - & 8.24 $\pm$ 1.34 \\
4215-02349-1 & - & +59 1923 & 1.03 $\pm$ 0.12 & 9.93 $\pm$ 0.16 & 1.59 $\pm$ 0.09 & - & 8.85 $\pm$ 1.49 \\
4421-01222-1 & - & +68 933 & 1.13 $\pm$ 0.21 & 9.83 $\pm$ 0.24 & 1.69 $\pm$ 0.16 & - & 10.41 $\pm$ 2.75 \\
4421-01437-1 & - & +68 937 & 1.16 $\pm$ 0.20 & 9.78 $\pm$ 0.24 & 1.42 $\pm$ 0.17 & - & 7.15 $\pm$ 1.93 \\
4421-01706-1 & - & +68 944 & 1.27 $\pm$ 0.23 & 9.66 $\pm$ 0.24 & 1.57 $\pm$ 0.17 & - & 8.79 $\pm$ 2.22 \\
4421-01779-1 & 160723 & +69 930 & 1.07 $\pm$ 0.14 & 9.88 $\pm$ 0.19 & 1.15 $\pm$ 0.14 & - & 5.19 $\pm$ 1.17 \\
4421-02304-1 & - & +68 931 & 1.08 $\pm$ 0.14 & 9.94 $\pm$ 0.17 & 1.68 $\pm$ 0.11 & - & 9.42 $\pm$ 1.95 \\
4421-02783-1 & - & +67 1023 & 1.17 $\pm$ 0.19 & 9.76 $\pm$ 0.23 & 1.48 $\pm$ 0.14 & - & 7.94 $\pm$ 1.73 \\
4421-02880-1 & 159966 & +68 938 & 1.22 $\pm$ 0.08 & 9.68 $\pm$ 0.10 & 1.68 $\pm$ 0.10 & 1.67 $\pm$ 0.09 & 10.51 $\pm$ 1.30 \\
4428-00192-1 & - & +67 1033 & 1.30 $\pm$ 0.22 & 9.66 $\pm$ 0.23 & 1.53 $\pm$ 0.17 & - & 8.59 $\pm$ 2.15 \\
4428-00560-1 & - & +68 958 & 1.28 $\pm$ 0.35 & 9.66 $\pm$ 0.32 & 2.05 $\pm$ 0.19 & - & 15.79 $\pm$ 5.32 \\
4428-01506-1 & - & +68 953 & 1.22 $\pm$ 0.13 & 9.74 $\pm$ 0.15 & 1.31 $\pm$ 0.10 & - & 6.79 $\pm$ 0.93 \\
4428-01561-1 & - & +68 951A & 1.19 $\pm$ 0.20 & 9.74 $\pm$ 0.23 & 1.46 $\pm$ 0.15 & - & 7.78 $\pm$ 1.73 \\
4428-01582-1 & - & +69 935 & 1.09 $\pm$ 0.12 & 9.93 $\pm$ 0.16 & 0.73 $\pm$ 0.17 & - & 3.11 $\pm$ 0.91 \\
4444-00200-1 & - & +68 1063 & 1.20 $\pm$ 0.12 & 9.81 $\pm$ 0.15 & 0.93 $\pm$ 0.07 & - & 4.24 $\pm$ 0.48 \\
4444-00717-1 & 184737 & +68 1071 & 0.93 $\pm$ 0.04 & 10.06 $\pm$ 0.06 & 1.97 $\pm$ 0.09 & 1.67 $\pm$ 0.14 & 10.46 $\pm$ 2.13 \\
4444-01116-1 & - & - & 1.18 $\pm$ 0.15 & 9.81 $\pm$ 0.19 & 1.55 $\pm$ 0.10 & - & 9.25 $\pm$ 1.53 \\
4445-00579-1 & 187178 & +68 1078 & 1.31 $\pm$ 0.19 & 9.59 $\pm$ 0.20 & 1.53 $\pm$ 0.18 & 1.62 $\pm$ 0.26 & 8.87 $\pm$ 2.66 \\
4448-00021-1 & 184873 & +69 1052 & 1.89 $\pm$ 0.12 & 9.09 $\pm$ 0.08 & 1.02 $\pm$ 0.20 & 1.66 $\pm$ 0.16 & 6.97 $\pm$ 1.63 \\
4448-00811-1 & - & +70 1068 & 1.29 $\pm$ 0.23 & 9.63 $\pm$ 0.22 & 1.44 $\pm$ 0.15 & - & 7.60 $\pm$ 1.55 \\
4448-01402-1 & - & - & 1.10 $\pm$ 0.13 & 9.93 $\pm$ 0.16 & 1.49 $\pm$ 0.09 & - & 8.39 $\pm$ 1.43 \\
4448-01430-1 & - & - & 1.22 $\pm$ 0.17 & 9.82 $\pm$ 0.19 & 1.49 $\pm$ 0.10 & - & 9.02 $\pm$ 1.53 \\
4448-01464-1 & - & - & 1.10 $\pm$ 0.10 & 9.90 $\pm$ 0.14 & 1.17 $\pm$ 0.09 & - & 5.66 $\pm$ 0.76 \\
4449-00492-1 & - & +69 1059 & 2.09 $\pm$ 0.20 & 9.08 $\pm$ 0.12 & 1.63 $\pm$ 0.10 & - & 8.48 $\pm$ 1.47 \\
4449-01168-1 & - & +70 1082 & 1.07 $\pm$ 0.08 & 10.01 $\pm$ 0.11 & 0.84 $\pm$ 0.12 & - & 3.68 $\pm$ 0.73 \\
4449-01170-1 & - & +69 1062 & 1.25 $\pm$ 0.19 & 9.66 $\pm$ 0.22 & 1.39 $\pm$ 0.16 & - & 7.17 $\pm$ 1.60 \\
4449-01543-1 & - & +69 1061 & 1.55 $\pm$ 0.19 & 9.39 $\pm$ 0.15 & 1.09 $\pm$ 0.13 & - & 4.65 $\pm$ 0.85 \\
4449-01785-1 & - & - & 1.03 $\pm$ 0.08 & 9.94 $\pm$ 0.11 & 0.78 $\pm$ 0.08 & - & 3.31 $\pm$ 0.42 \\

\hline 
\multirow{2}{*}{0435-03989-1} &\multirow{2}{*}{-}&\multirow{2}{*}{-}&2.02 $\pm$ 0.11 &9.09 $\pm$ 0.07 &1.60 $\pm$ 0.06&	-				& 7.97 $\pm$ 0.96\\
  				  								&&	&1.37 $\pm$ 0.16 &9.52 $\pm$ 0.15 &1.25 $\pm$ 0.12&	-				& 5.90 $\pm$ 1.03\\
\multirow{2}{*}{1058-01329-1} &\multirow{2}{*}{}&\multirow{2}{*}{+08 4256}&1.37 $\pm$ 0.13 &9.51 $\pm$ 0.12 &1.33 $\pm$ 0.09&	-				& 6.55 $\pm$ 0.86\\
				  								&&	&1.90 $\pm$ 0.09 &9.15 $\pm$ 0.05 &1.58 $\pm$ 0.06&	-				& 8.19 $\pm$ 0.82\\
\multirow{2}{*}{1208-00699-1} &\multirow{2}{*}{-}&\multirow{2}{*}{-}&1.64 $\pm$ 0.22 &9.34 $\pm$ 0.15 &1.89 $\pm$ 0.12&	-				&13.89 $\pm$ 2.47\\
				 								&&	 &1.13 $\pm$ 0.11 &9.83 $\pm$ 0.14 &1.64 $\pm$ 0.10&	-				&11.03 $\pm$ 1.73\\
\multirow{2}{*}{1423-00270-1} &\multirow{2}{*}{-}&\multirow{2}{*}{+20 2473}&1.01 $\pm$ 0.10 &9.86 $\pm$ 0.15 &1.65 $\pm$ 0.08&	-				& 9.58 $\pm$ 1.23\\
												&&	  &1.27 $\pm$ 0.07 &9.52 $\pm$ 0.07 &1.84 $\pm$ 0.05&	-				&11.38 $\pm$ 1.07\\
\multirow{2}{*}{2822-00812-1} &\multirow{2}{*}{-}&\multirow{2}{*}{+40 303}&1.58 $\pm$ 0.18 &9.38 $\pm$ 0.13 &1.80 $\pm$ 0.10&	-				&12.36 $\pm$ 2.02\\
				 								&&	 &1.13 $\pm$ 0.12 &9.84 $\pm$ 0.15 &1.59 $\pm$ 0.08&	-				&10.10 $\pm$ 1.48\\
\multirow{2}{*}{3105-00152-1} &\multirow{2}{*}{-}&\multirow{2}{*}{-}&1.22 $\pm$ 0.08 &9.68 $\pm$ 0.10 &1.66 $\pm$ 0.04&	-				&10.62 $\pm$ 0.83\\
												&&	  &1.48 $\pm$ 0.09 &9.41 $\pm$ 0.08 &1.80 $\pm$ 0.05&	-				&12.08 $\pm$ 0.99\\
\multirow{2}{*}{3122-02192-1} &\multirow{2}{*}{-}&\multirow{2}{*}{-}&1.26 $\pm$ 0.06 &9.66 $\pm$ 0.07 &0.69 $\pm$ 0.06&	-				& 2.87 $\pm$ 0.29\\
												&&	  &1.02 $\pm$ 0.06 &9.95 $\pm$ 0.08 &0.57 $\pm$ 0.07&	-				& 2.54 $\pm$ 0.29\\
\multirow{2}{*}{3227-00413-1} &\multirow{2}{*}{216536}&\multirow{2}{*}{+43 4329}&1.70 $\pm$ 0.34 &9.27 $\pm$ 0.20 &1.99 $\pm$ 0.17&	-				&14.77 $\pm$ 4.09\\
												&&	  &1.11 $\pm$ 0.11 &9.83 $\pm$ 0.16 &1.66 $\pm$ 0.10&	-				&10.96 $\pm$ 2.36\\
\multirow{2}{*}{3319-00892-1} &\multirow{2}{*}{20076}&\multirow{2}{*}{+48 873}&1.70 $\pm$ 0.15 &9.27 $\pm$ 0.10 &1.70 $\pm$ 0.07&	-				& 9.88 $\pm$ 1.19\\
												&&	  &1.33 $\pm$ 0.11 &9.55 $\pm$ 0.12 &1.50 $\pm$ 0.09&	-				& 8.29 $\pm$ 1.10\\
\multirow{2}{*}{3805-00709-1} &\multirow{2}{*}{77819}&\multirow{2}{*}{+53 1309}&1.30 $\pm$ 0.05 &9.55 $\pm$ 0.05 &1.09 $\pm$ 0.08&   \multirow{2}{*}{1.52$\pm$0.19} 	& 6.28 $\pm$ 1.28\\
												&&	  &1.90 $\pm$ 0.17 &9.13 $\pm$ 0.10 &1.35 $\pm$ 0.15&        				& 6.81 $\pm$ 1.44\\
\multirow{2}{*}{4211-00364-1} &\multirow{2}{*}{-}&\multirow{2}{*}{+67 1020}&1.27 $\pm$ 0.10 &9.62 $\pm$ 0.11 &1.48 $\pm$ 0.06&	-				& 8.06 $\pm$ 0.76\\
												&&	  &1.72 $\pm$ 0.13 &9.26 $\pm$ 0.09 &1.62 $\pm$ 0.07& 	-				& 9.47 $\pm$ 0.94\\
\hline
\end{longtable}
\tablefoot{The following columns represent: (1-3) identification from SIMBAD, (4-8) astrophysical stellar parameters and their uncertainties (mass, age, luminosity from this work, luminosity calculated from parallax (if available) radius).}
\end{longtab}

\end{document}